\documentclass{article}

\usepackage{arxiv}

\usepackage[utf8]{inputenc} % allow utf-8 input
\usepackage[T1]{fontenc}    % use 8-bit T1 fonts
\usepackage{hyperref}       % hyperlinks
\usepackage{url}            % simple URL typesetting
\usepackage{booktabs}       % professional-quality tables
\usepackage{amsfonts}       % blackboard math symbols
\usepackage{nicefrac}       % compact symbols for 1/2, etc.
\usepackage{microtype}      % microtypography
\usepackage{lipsum}
\usepackage{graphicx}
\usepackage{natbib} 
\graphicspath{ {./images/} }

\usepackage{newtxtext,newtxmath}
\newcommand{\dif}{\mbox{d}}

\title{THE STRATIFICATION OF REGOLITH ON CELESTIAL OBJECTS}

\author{
Rainer Schr\"apler \\
 Institut f\"ur Geophysik und extraterrestrische Physik\\
 University of Braunschweig\\
 Mendelssohnstr. 3, D-38106 Braunschweig\\
  \texttt{r.schraepler@tu-braunschweig.de} \\
 \And
  J\"urgen Blum \\
  Institut f\"ur Geophysik und extraterrestrische Physik\\
  University of Braunschweig\\
  Mendelssohnstr. 3, D-38106 Braunschweig\\
  \And
  Ingo von Borstel \\
  Institut f\"ur Geophysik und extraterrestrische Physik\\
  University of Braunschweig\\
  Mendelssohnstr. 3, D-38106 Braunschweig\\
  \And
  Carsten G\"uttler \\
  Institut f\"ur Planetologie\\
  University of M\"unster\\
  Wilhelm-Klemm-Str. 10, D-48149 Münster, Germany\\
  }

\begin{document}

\maketitle
%\makeauthor

%\title{THE STRATIFICATION OF REGOLITH ON CELESTIAL OBJECTS}

%\shorttitle{REGOLITH STRATIFICATION}
%\author{\scshape Rainer Schr\"apler,  J\"urgen Blum and Ingo von Borstel}
%\affil{Institut f\"ur Geophysik und extraterrestrische Physik, University of Braunschweig \\
%Mendelssohnstr. 3, D-38106 Braunschweig, Germany}
%\email{r.schraepler@tu-bs.de}
%\and
%\author{\scshape Carsten G\"uttler}
%\affil{Max Planck Institute for Solar System Research\\
%Justus-von-Liebig-Weg 3, D-37077 G\"ottingen, Germany}

%\shortauthors{Schr\"apler, Blum, von Borstel \& G\"uttler}

%\author[IGEP]{Rainer Schr\"apler}
%\ead{r.schraepler@tu-bs.de}
%\author[IGEP]{J\"urgen Blum}
%\author[IGEP]{Ingo von Borstel}
%\author[MPS,IGEP]{Carsten G\"uttler}
%\address[IGEP]{Institut f\"ur Geophysik und extraterrestrische Physik, University of Braunschweig Mendelssohnstr. 3, D-38106 Braunschweig, Germany}
%\address[MPS]{Max Planck Institute for Solar System Research, Justus-von-Liebig-Weg 3, D-37077 G\"ottingen, Germany}

\begin{abstract}
All atmosphere-less planetary bodies are covered with a dust layer, the so-called regolith, which determines the optical, mechanical and thermal properties of their surface. These properties depend on the regolith material, the size distribution of the particles it consists of, and the porosity to which these particles are packed. We performed experiments in parabolic flights to determine the gravity dependency of the packing density of regolith for solid-particle sizes of 60~$\mu$m and 1~mm as well as for 100-250~$\mu$m-sized agglomerates of 1.5~$\mu$m-sized solid grains. We utilized g-levels between 0.7~m~s$^{-2}$ and 18~m~s$^{-2}$ and completed our measurements with experiments under normal gravity conditions. Based on previous experimental and theoretical literature and supported by our new experiments, we developed an analytical model to calculate the regolith stratification of celestial rocky and icy bodies and estimated the mechanical yields of the regolith under the weight of an astronaut and a spacecraft resting on these objects.
\end{abstract}

%\begin{keyword}
methods: laboratory  --- small bodies
%\end{keyword}
%\end{frontmatter}
\vspace*{0mm}

\section{INTRODUCTION}
\label{kap:INTRO}
\citet{Gary1972} state that ``Regolith is a mantle of loose incoherent rocky material of various origins that forms the surface of planetary bodies''. Its size distribution determines the optical, mechanical and thermal properties of these surfaces. On bodies without atmosphere, regolith is formed by high-velocity impacts of interplanetary particles of different sizes. These impacts produce ejecta material with a size and velocity distribution determined by the impactor and target properties. Ejecta with velocities lower than the escape velocity of the target body are reaccreted and, thus, form the regolith on the surface. Laboratory experiments of \citet{Hartmann1985} found ejecta velocities depending on the velocity of the impactor. The experiments of \citep[e.g.][]{Fujiwara1980,Nakamura1994} and studies of crater structures on the Moon \citep{Vickery1986,Vickery1987} showed that smaller particles are ejected at higher velocities than the larger ejecta. Finally \citet{AsteroidsIII} calculated in their chapter III the trajectories and re-accretion of ejected material on small celestial objects depending on its  velocity. Therefore, the regolith size distribution should depend on the escape speed and, hence, on the size of the target body. \citet{Gundlach2012} developed a method to correlate the size of the regolith particles with the thermal conductivity of the regolith. Thus, from measurements of the thermal inertia of small objects in the Solar System, their regolith-particle size can be estimated. The regolith filling factor, i.e. the packing density of the regolith particles (or 1 - porosity), enters their calculation as a free parameter. \citet{Gundlach2012} found that objects with a size smaller than 100 km are covered with regolith particles of the size of $\sim$1-10~mm, whereas larger objects carry regolith with particle sizes of 10-100~$\mu$m.

\citet{Skorov2012} model comet nuclei as consisting of macroscopic dust and ice agglomerates. In a recent work, \citet{Blumetal2014} showed that this model can explain the observed continuous activity of comets in the inner Solar System. Inside the comet nucleus, the dust and ice agglomerates are only subjected to the weak gravitational force of the comet, which leads to a stratification of the packing density within a comet nucleus.
On celestial objects the regolith is compacted by their gravitational force, which leads to a stratification where the filling factor of the regolith increases with layer depth.

However, the texture (e.g. particle size distribution and packing density) and vertical stratification of regolith is basically unknown for all celestial objects except for the Moon \citep{LunarSourcebook}. Recently, \citet{Kiuchi2014} related the particle size to the porosity for a variety of celestial objects. However, they used a simplified model in their analysis that implies that the packing density depends only on the ratio of the gravitational force of the contact force of a regolith particle. A result of this simplification is that the packing density of the regolith does not change with depth. If this was true, it would be impossible to compress loosely packed regolith at all. Thus, the results of \citet{Kiuchi2014} are valid for the uppermost layer of the regolith where hydrostatic compression is small.

We will show in this article that stratification of regolith due to hydrostatic compression is important and can be calculated with the relation between pressure and filling factor of the regolith found in our article. With this information, also the yield of objects on the surface of the regolith-covered celestial objects was derived.

For $\mu$m sized grains, this relation is experimentally obtained (see Section \ref{sect:EXACT} by compressing the regolith with a piston in a cylinder as shown by \citet{Guettleretal2009}. For the 10 $\mu$m grains, we measured the low-pressure part of the compression curve by adding thin layers on a regolith to compress it by its own weight and the high pressure part by compressing the regolith by a piston. For grains larger than 50 $\mu$m, the regolith is compacted to RCP (random close packing, its highest possible filling factor) by its own gravitational force for thicknesses above a few particle layers. We therefore did most of these measurements in the low-gravity environment aboard the Zero-G-plane of the European Space Agency. However, in spite of that, it was only possible to measure the upper 20\% of the compression curve (see Section \ref{sect:Expres}). It was therefore necessary to develop a model for the full compression curve (see Section \ref{sect:CE}) in which only one parameter must be experimentally determined. With this analytical model, we could additionally reproduce the relation between filling factor and grain radius found by \citet{Yangetal2000} at constant $g$ and can show the consistency of our model (see Sect. \ref{sect:CE}).

To simulate the stratification of the packing density within a comet nucleus, we additionally used in our experiments 0.1-0.25~mm-sized dust agglomerates consisting of 1.5~$\mu$m-sized mono-disperse and spherical $\rm SiO_2$ grains. We performed our measurements with mono-disperse spherical monomer particles to ease the theoretical approach to our measurements, whereas the particles on celestial objects will most likely be irregular and polydisperse. However, our previous measurements in \citet{BSDR2006} showed that the results will not change dramatically for these particles (see Sect. \ref{sect:appli}).

In Section \ref{sect:EXACT}, we describe our experimental approach and the experimental findings. Section \ref{sect:CE} uses the analytical-approximation form of \citet{Guettleretal2009} and the results of \citet{DoTi1995} and \citet{krijtetal2014} to  model the observed regolith stratification. In Section \ref{sect:appli}, we derive filling-factor profiles of the regolith for selected small rocky and icy bodies and estimate the mechanical yield of the regolith under the weight of an astronaut and a spacecraft, resting on these objects. Finally, Section \ref{kap:COCON} summarizes our results.

\section{EXPERIMENTAL APPROACH} \label{sect:EXACT}

In this Section, we describe the experimental methods developed and applied for the determination of the pressure-dependent filling factor of regolith and the results we received in parabolic-flight and laboratory experiments.

\subsection {Particle Samples}\label{sect:EP}
\begin{figure}[!tb]
    \center
      \includegraphics[width=0.45\textwidth]{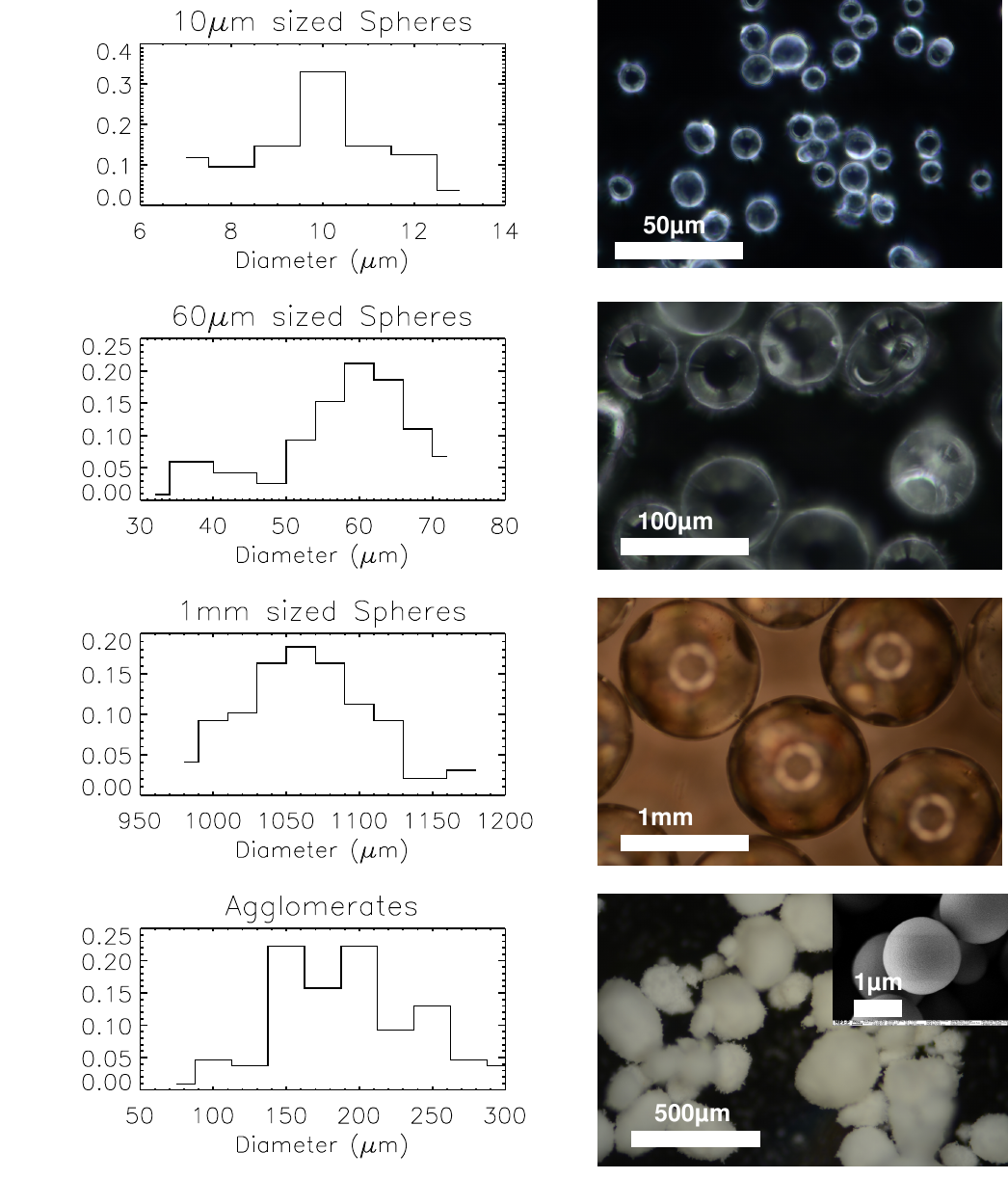}%\columnwidth
    \caption{\label{fig:Histo}
    Measured number-fraction histograms of the diameters of the regolith-analog particles (left column) and the corresponding microscopy images of our particle species. The inset in the bottom-right picture shows the constituent 1.5~$\rm \mu m$-sized $\rm SiO_2$ grains of the dust agglomerates.}
\end{figure}

As regolith analogs, we used amorphous SiO$_2$ spheres with diameter of 10~$\mu$m, glass spheres with 60~$\mu$m diameter from unknown glass type, spherical soda lime particles with 1 mm diameter, and agglomerates of 1.5~$\mu$m-sized amorphous SiO$_2$ spheres with 100-250~$\mu$m diameter.
Fig. \ref{fig:Histo} shows the size distributions and microscopic images
of these samples.
The experiments were performed at 10$^5$ Pa ambient pressure. Laboratory, microgravity and hypergravity experiments were performed with all samples with the exception of the 10~$\mu$m-sized particles. These particles fluidized under microgravity conditions, due to the ambient gas pressure, and could, thus, not deliver reliable results. For these particles, we only analyzed the experiments conducted in the laboratory. The humidity of the air inside the experiment could increase the contact forces of the particles and therefore decrease their filling factor at a given pressure. To find this influence we did ground experiments for the 10~$\mu$m-sized particles at $10$ Pa gas pressure and at ambient (10$^5$ Pa) air pressure (asterisks and triangles in Fig. \ref{fig:PPress}) and found no deviation within our error bars.

\subsection {Experimental Setup}\label{sect:ESPL}
\subsubsection{Parabolic-Flight Experiments}\label{sect:ESPL_PFE}
Parabolic-flight experiments under reduced and hyper-gravity conditions were performed onboard the ZERO-G Airbus A300 aircraft. Typically, several experimental setups are mounted in the passenger area of the aircraft and are operated by onboard experimentalists during the flights. In one such flight, the aircraft performs up to 31 parabolic and catenary-curve flight maneuvers.

During a parabolic flight maneuver, inertia forces completely cancel Earth's gravity and the aircraft and all experiments onboard are completely weightless. During a catenary-curve flight maneuver, inertia forces partly cancel Earth's gravity and the aircraft and all experiments are partly weightless so that the environment of celestial bodies with surface accelerations smaller than on Earth can be simulated.

In the flights described here, the flight maneuvers consisted of 13 catenary curves with Martian (0.38 $g$), 12 with Lunar (0.17 $g$) and 6 parabolas at zero gravity, respectively. Here $g=9.81~\rm m~s^{-2}$ is the surface acceleration on Earth. As the aircraft has to enter and exit the parabolas and catenary curves, a "pull-up" and a "pull-out" maneuver are required before and after each constant-acceleration curve. These maneuvers cause hyper-gravity accelerations from 1.3 $g$ to 2 $g$. We also utilized these hyper-g phases for our experiments.

In the parabolic-flight experiments considered here, four different masses of otherwise identical regolith samples were filled in glass cylinders with an inner diameter of 2.5 cm, with cylinder no. 1 having the highest mass and cylinder no. 4 with the lowest mass, respectively.

Even for the largest (1 mm) particles, only 1\% of their contact points are with cylinder walls so that the finite cylinder diameter will influence the measurements only through the Janssen effect \citep{Janssen1895,Sperl2005} described below and quantified in Eq. \ref{eq:jan}. It should also be noted that the lowermost layer of particles naturally possesses a different filling factor, because of the flat surface of the cylinder base. This boundary effect is relevant for the mm-size particle measurements and was corrected for in the data analysis.

\begin{figure}[!tbh]
    \center
      \includegraphics[width=.45\textwidth]{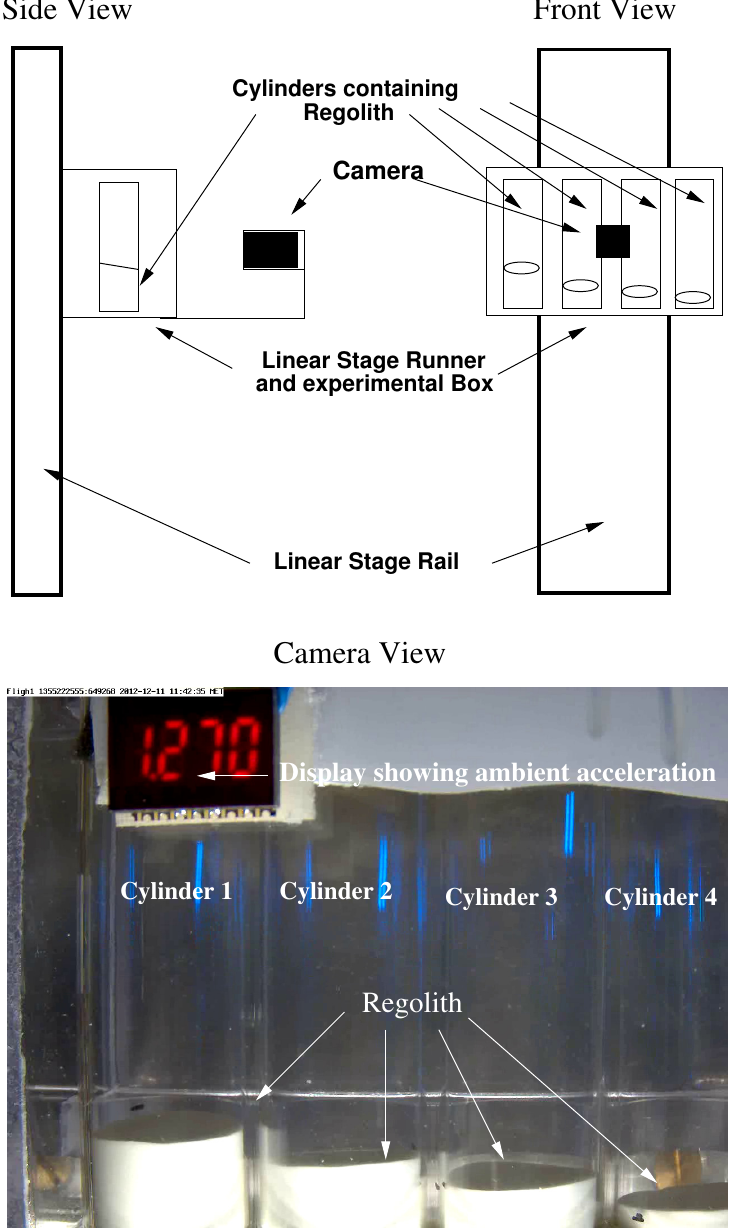}%\columnwidth
    \caption{\label{fig:Pic} 
    Top: Schematic drawing of the experimental setup. Bottom: Picture of the experimental box with the four sample cylinders containing regolith from 60 $\mu$m-sized spheres, captured with the video camera. The LED display on the upper left of the image shows the acceleration level inside the airplane.}
\end{figure}

The four glass cylinders were placed in a polycarbonate experimental box and mounted on a linear stage together with a video camera (see Fig. \ref{fig:Pic}). The video camera was used to measure the filling height of each of the samples. Caused by the mounting of the cylinders in the polycarbonate box, their bottom was not visible in the field of view of the camera (see Fig. \ref{fig:Pic}). Therefore, we calibrated the filling-height by comparing the measurements during the 1~$g$ phases with height measurements on the ground without the box.

During the parabolic flights, we performed two types of experiments.
\begin{itemize}
  \item Type-1 experiments: The regolith is deagglomerated and homogeneously distributed inside the glass cylinder by a single shake using a succession of an upward 20~$g$ acceleration, a downward -20~$g$ acceleration, and then again an upward 20~$g$ acceleration so that in the end the sample box comes to rest at its starting position. This initial shaking with an amplitude of 20 cm was performed using the electromagnetic linear stage. Thereafter, the regolith particles sedimented in the ambient (log-$g$ or hyper-$g$) acceleration downwards and formed a layer whose filling height could be measured on the video-camera images (see Fig. \ref{fig:Pic}). During this sedimentation phase, the residual-acceleration vector in the aircraft is not perfectly directed in vertical direction. Therefore, surfaces of the regolith in the 4 cylinders are slightly tilted (in all 4 cylinders in the approximate same direction and with the same angle). We only used measurements in which the surface was tilted towards the camera so that the full perimeter of the sample surface was visible. We measured the fill height at four points at the cylinder rim, that have a distance of 90 degrees and calculated the mean fill height by averaging over these four measurements. A random deviation of the surface from a perfect plane is a reason for the  large error bars in our figures \ref{fig:DFF}, \ref{fig:PPress} and \ref{fig:PoR}, because an error in the fill height measurements of $\pm$0.3 mm corresponds to an error in $\Phi$ of $\pm$0.1.  The error bars of this measurements are found statistically by calculating the mean and standard deviation of different measurements.

      The ambient acceleration level is produced by the airplane that performs catenary-curves with Martian, Lunar and hyper-gravity accelerations. As the acceleration level provided by the aircraft varied strongly with time, particularly during the hyper-gravity phases, and as the experiment duration (i.e., shaking plus sedimentation) was always short compared to these variations, we were able to measure the fill heights for about 50 uniformly distributed $g$-levels between $\sim0.3~g$ and $\sim2~g$ during the parabolic-flight campaign. For ease of analysis, we equipped our experiment with an acceleration sensor whose read-out value (in units of $g$) was displayed in the field of view of the video camera (see the LED display in Fig. \ref{fig:Pic}).

  \item Type-2 experiments: These experiments were performed at micro-gravity conditions generated by parabola maneuvers of the airplane. The regolith is deagglomerated and homogeneously distributed inside the glass cylinder as described in the type-1 experiments. In a subsequent second step, the linear stage is very slowly (at 5 cm s$^{-1}$) moved upwards to bring all regolith particles back to the base of the cylinder. In a third step, the linear stage accelerated the experimental box upwards to compress the granular medium at pre-defined constant acceleration levels of 0.05, 0.1, 0.15 and 0.2~$g$, respectively, over a length of 71 cm, causing acceleration times of 3.7 , 3.1 and 2.7~s, respectively. Naturally, these accelerations were superposed with the residual accelerations of the aircraft, which were in the range of $\pm$ 0.2~$g$. We were able to perform up to three type-2 experiments per parabola. Because the temporal changes of the residual acceleration were long compared to the experimental duration, we again obtained a total of around 40 equally distributed $g$-levels between $\sim0.06~g$ and $\sim 0.3~g$.

\end{itemize}

\subsubsection {Ground Experiments}\label{sect:LE}
For the experiments exclusively performed in the laboratory under normal-gravity conditions, we successively filled a small mass $\dif m$ of regolith simulant particles in a glass tube with 25 mm diameter and measured the resulting gain in volume, $\dif V$. The ratio $\dif m / \dif V$ is the density of the lowermost level in the glass tube. Repeating this procedure by carefully stratifying additional layers with a spatula resulted in a filling-factor profile along the tube height (type 1 ground experiments).

As it was not possible to perform reduced-gravity experiments for the 10 $\mu$m sized particles during the parabolic flights, we additionally conducted type 2 ground experiments. The difference to the regular (type 1) ground experiments was that we additionally shook the regolith before measuring the fill height to avoid any hysteresis effect. To obtain high pressure data of the 10 $\mu$m sample, we also performed compression measurements as described in \citet{Guettleretal2009} (type 3 ground experiments). We performed these experiments in vacuum at 10 Pa gas pressure (triangles in Fig. \ref{fig:PPress}) and at ambient air pressure (10$^5Pa$; asterisks in Fig. \ref{fig:DFF}) and found that there is no difference within our error bars.

An advantage of using 10 $\mu$m sized particles over larger grains is that the hydrostatic compression by their own weight is much smaller than the pressure required to compress them to RCP. Therefore, higher hydrostatic pressures were simulated by compressing the regolith with a piston. Thus, we could achieve measurements over a wide range of the compression curve for the 10 $\mu$m sized particles. For the 60 $\mu$m and 1 mm particles, only segments of this curve could be obtained (see Fig. \ref{fig:PPress}).

The pressure acting within the regolith bottom layer in the cylinder scales with the fill height $x$ of the cylinders following Janssen's Equation \citep{Janssen1895,Sperl2005}
\begin{eqnarray}
p=\frac {m g}{4 K \pi r^2}[1-\mbox{exp}(-4K\frac{x}{2 r})],\label{eq:jan}
\end{eqnarray}
where $K=0.2$ \citep{Janssen1895}, $m$, and $r$ are the ratio between the horizontal and vertical components of the stress tensor in the regolith, the mass inside the cylinder at a given fill height $x$, and the radius of the cylinder, respectively. As \citet{Janssen1895} used containers with a square cross section but we employed circular cylinders instead, we substituted the side length by $2 r$.

\subsection{Experimental Results}\label{sect:Expres}

\subsubsection{Parabolic-Flight Experiments}\label{sect:PE}
As shown in Sect. \ref{sect:ESPL_PFE}, our measurements in the parabolic-flight experiments directly yielded the $g$-dependence of the average filling factor of the regolith in each of the four glass tubes. As these four glass tubes were always filled with the same sample material but to different heights, we used the method depicted in Sect. \ref{sect:LE} to derive the height-dependent filling factor. Thus, cylinder 4 with the smallest fill height corresponds to the uppermost layer of the other three cylinders and possesses an average filling factor of
\begin{eqnarray}
\Phi_4=\frac{m_4}{V_4 \rho_b},
\end{eqnarray}
where m$_4$, $V{_4}$, and $ \rho_b$ are the mass and volume of the regolith in cylinder 4, and the density of the regolith bulk material, respectively. The filling factor of the subsequent layers $i=3, 2, 1$ can then be calculated by
\begin{eqnarray}
\Phi_i=\frac{m_{i}-m_{i+1}}{(V_{i}-V_{i+1}) \rho_b},
\end{eqnarray}
with $m_i$ and $V_i$ being the mass and volume of cylinder $i$, respectively.

\begin{figure}[!tb]
    \center
      \includegraphics[width=0.45\textwidth]{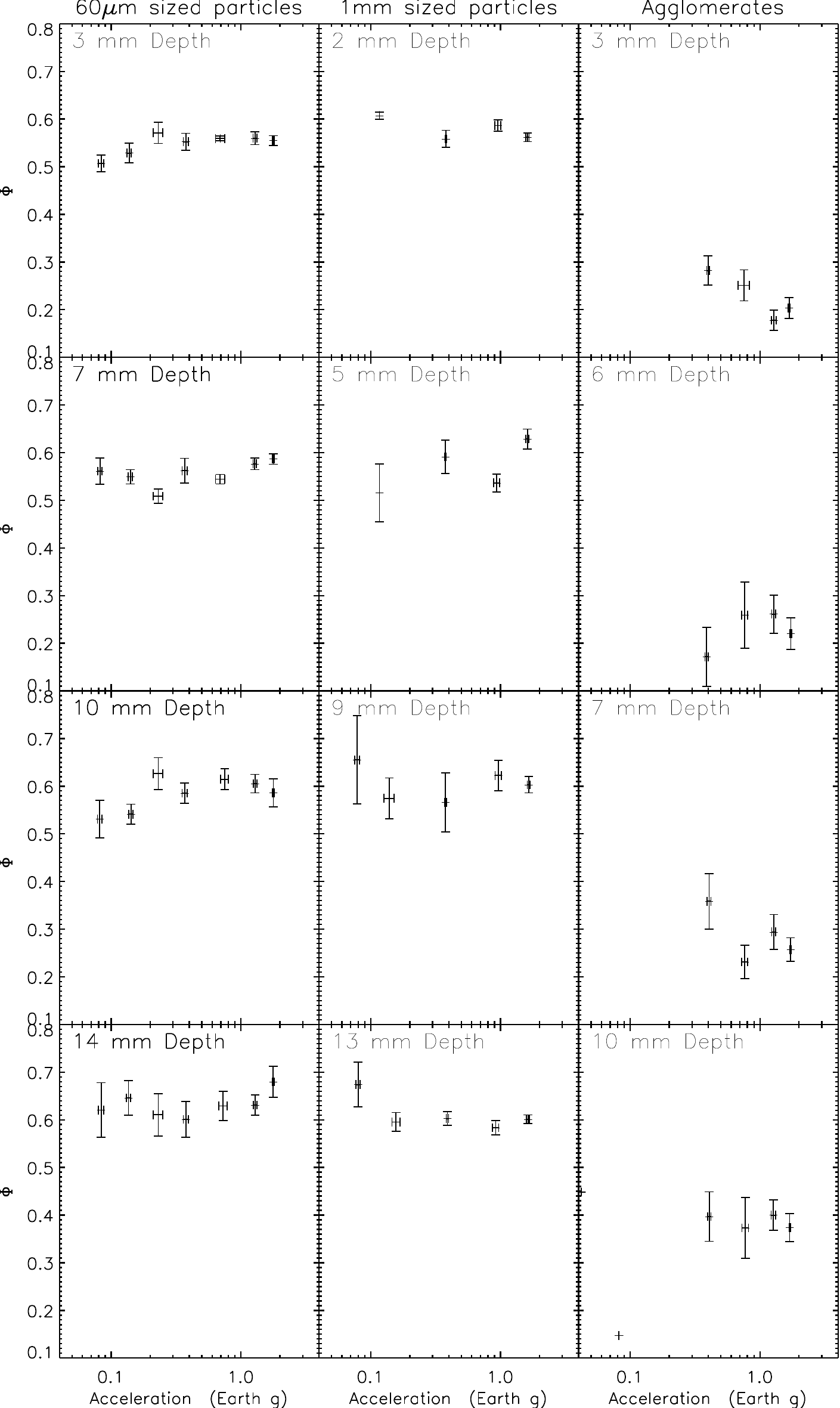}%\columnwidth
    \caption{\label{fig:DFF} The filling factor $\Phi$ for regolith layers consisting of 60~$\mu$m-sized spheres (left), 1~mm-sized spheres (middle), and 100-250~$\mu$m-sized dust agglomerates (right), respectively, as a function of the ambient acceleration level (in units of $g=9.8$~m~s$^{-2}$) at depths of 3~mm, 7~mm, 10~mm and, 14~mm (from top to bottom; left), 2~mm, 5~mm, 9~mm and, 13~mm (middle), and 3~mm, 6~mm, 7~mm and, 10~mm (right), respectively. Each data point in the graph represents the mean of 10 - 12 (left), 16 - 20 (middle), and 17 - 21 (right) individual measurements, respectively, and the error bars denote one standard derivation.}
\end{figure}

Figs. \ref{fig:DFF} shows the resulting filling factors $\Phi_i$ ($i=4, 3, 2, 1$ from top to bottom) of the 	%RAUS DAMITsimulated
regolith layers consisting of 60~$\mu$m-sized particles, 1~mm-sized particles, and 100-250~$\mu$m-sized dust agglomerates, respectively, as a function of the ambient acceleration level. The error bars denote the standard derivation among the 10-12 (60~$\mu$m-sized particles), 16-20 (1~mm-sized particles), and 17-21 measurements (100-250~$\mu$m-sized dust agglomerates), respectively.

\begin{figure*}[!tb]
    \center
      \includegraphics[width=1.\textwidth]{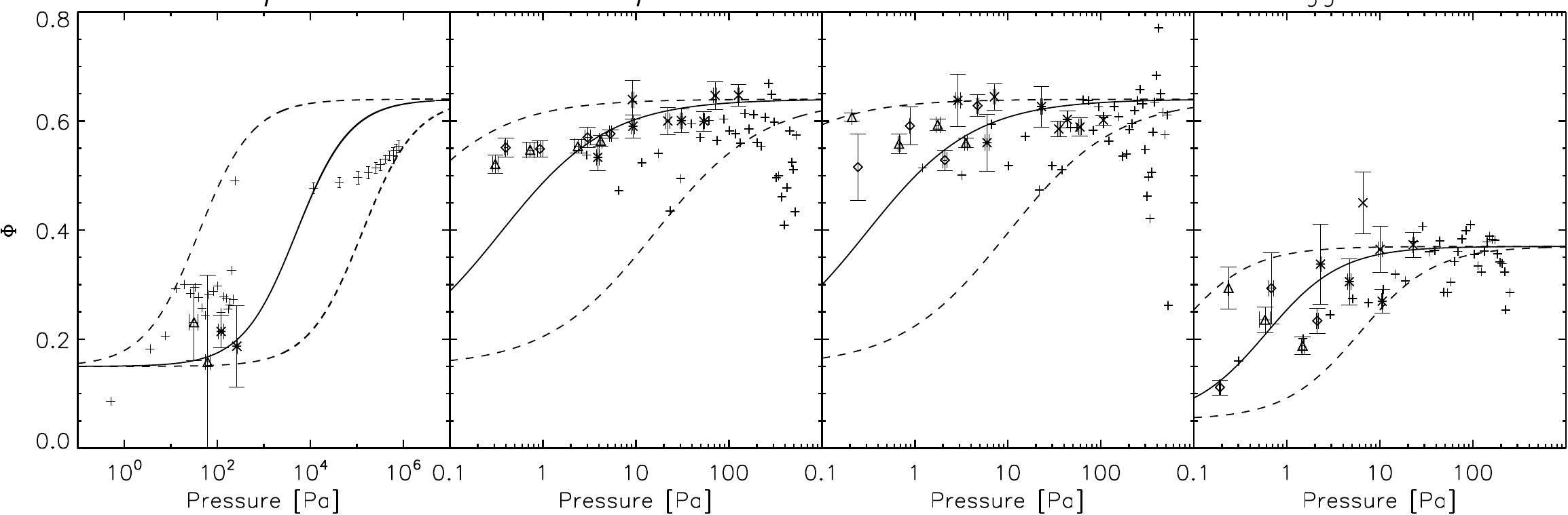}%\columnwidth
    \caption{\label{fig:PPress} The filling factor $\Phi$ for regolith layers consisting of (from left to right) 10~$\mu$m-sized, 60~$\mu$m-sized, and 1~mm-sized spheres as well as for 100-250~$\mu$m-sized dust agglomerates as a function of the hydrostatic pressure, following Eq. \ref{eq:jan}. This pressure was calculated from the layer depth and the ambient acceleration level as measured in the experiments. Triangles represent data points from the uppermost layer, diamonds from the second layer, asterisks form the third layer, and crosses from the bottom layer (for the 60~$\mu$m and 1~mm-sized spheres and for the 100-250~$\mu$m-sized dust agglomerates only). These data points and their error bars were measured in the parabolic-flight experiments and are identical to those shown in Fig. \ref{fig:DFF}. Pluses represent boxcar averages of the filling factor measured in the type 1 ground experiments. Mind that for the 10~$\mu$m-sized particles, only ground-based data are available. For these particles, the data measured in type 2 ground experiments are shown by triangles (vacuum) an asterisks (air) and for the type 3 ground experiments by pluses with error bars. The solid line corresponds to the model curve as given by Eq. \ref{eq:gu}. It is fitted to our measurements using $p_m$  as a fit parameter. The upper and lower dashed lines correspond to Eq. \ref{eq:gu} with $p_{\rm min}$ and $p_{\rm max}$, respectively. Mind the different pressure scale for the 10~$\mu$m-sized particles.}
\end{figure*}

The graphs for the 60 $\mu$m particles generally show an increase of the filling factor with increasing depth and increasing gravity level, until RCP  ($\Phi_{\mathrm{RCP}}$=0.64) is reached. The filling factor for the mm-sized particles is close to RCP for the whole gravity and depth ranges. As the individual measurement errors are very large, it is difficult to recognize a systematic behavior for these particles. The interpretation for this data is easier in Fig. \ref{fig:PPress}, because there the data of layer depth and gravity level are combined to give a better statistical basis. The dust agglomerates are, however, only compressed to a maximum filling factor of $\Phi = 0.37$ and never get close to RCP.

In all our measurements, the regolith particles adhered in mono-layers to the walls and to the top of the cylinders. However, this considerably affects the resulting filling factors only for the mm-sized particles. We could partly correct for this effect for the 1~mm-sized spheres by counting the (spatially resolved) particles on the walls, but not all surfaces were visible in the camera field of view. For the invisible surfaces, we assumed the same surface coverage of adhered particles as on the visible surfaces. However, we have to bear in mind that this procedure might lead to an overcorrection of the filling factor, because the invisible surfaces were in the upper part of the cylinders and, thus, not necessarily came in contact with the same number of particles.

The effect of adhesion is stronger at low $g$-levels and leads to the artificial decrease (because of the overcorrection described above) of the filling factor for accelerations between $0.1~g$ and $0.3~g$ (see Fig. \ref{fig:DFF}), which is visible in the data of the 1~mm-sized particles. These particles also show a boundary effect in cylinder 4, because of the rather small layer thickness of 4~mm and a naturally reduced filling factor in the first particle layer from the cylinder bottom.

It was not possible to use 10 $\mu$m sized particles in the parabolic-flight experiments. The inside of the cylinders were at ambient pressure and an air cushion formed in between the particles. The time available for sedimentation during the experiment was too short for the air to diffuse out of the regolith layers. This effect also led to a systematic reduction of the filling factor for the 60 $\mu$m particle measurements at low pressures.

However, the model described in Section \ref{sect:CE} will show that the filling factor only depends on the hydrodynamic pressure inside the regolith. Therefore, in Fig. \ref{fig:PPress}, we plotted all reduced- and hyper-gravity data points as a function of the respective hydrostatic pressure inside the regolith, using Eq. (\ref{eq:jan}), together with the data from the ground experiments (see Sect. \ref{sect:LES}).

\subsubsection{Ground Experiment}\label{sect:LES}
Microscopically, restructuring and compaction inside a regolith layer takes place, because external forces act against rolling friction on inter-particle contact points \citep{Yangetal2000}. These forces arise from the ambient pressure inside the regolith and their origin (gravity-driven hydrostatic or external by a piston) is irrelevant. Therefore, the compression of regolith can also be studied with ground-based experiments for pressures exceeding the hydrostatic value. The results of our type 1 ground experiments, as described in Sect. \ref{sect:LE}, are shown as pluses in Fig. \ref{fig:PPress} and were derived by boxcar-averaging of four individual filling-factor measurements as a function of hydrostatic pressure. We chose to use a boxcar-averaging method instead of binning the data points (which was done for the parabolic-flight experiments), because the experiments showed a hysteresis effect, which would have been veiled by the binning method. Between two subsequent measurements in the laboratory, we added a defined mass of particles to the sample (see Sect. \ref {sect:LE}).

In the type 2 ground experiments measurements, the regolith was shaken up after a new upper layer had been deposited so that all regolith particles could sediment simultaneously. With this, we got rid of the above-described hysteresis effect and could bin the data for better statistics. These data (only available for the 10 $\mu$ m particles) are shown in Fig. \ref{fig:PPress} as triangles (measurements in vacuum) and asterisks (measurements at 10$^5$ Pa air) with error bars. In the type 3 ground experiments, we produced high pressures by compressing the regolith with a piston as described in \citet{Guettleretal2009}. The results of these experiments (only for 10 $\mu$ m-particles) are shown as pluses with error bars in Fig. \ref{fig:PPress}. Here, each error bar denotes the standard deviation of four different measurements. These measurements start at around 10$^4$ Pa pressure. As can be seen in Fig. \ref{fig:PPress}, slightly increasing pressure does not appreciably compact the regolith. We think that is because the internal friction of our compression apparatus has to be overcome. At around 10$^5$ Pa pressure, the regolith then gets compacted. At 10$^6$ Pa pressure, our measurements terminate, because the limit of our experimental apparatus had been reached.

Fig. \ref{fig:PPress} shows a general increase of the filling factor with increasing pressure, but in some intervals of the type 1 ground experiments, the filling factor decreases with increasing pressure. This is due to the fact that a granular medium does not always restructure under the weight of the additional layers, which then leads to a reduction in filling factor. After the additional layers reach a height between 4~mm and 7~mm, corresponding to an additional pressure of 20 to 35~Pa, the regolith suddenly restructures to a higher filling factor. The effect is exaggerated in the data shown in Fig. \ref{fig:PPress}, because the hysteresis, which takes place over the whole cylinder height, is assigned to only one layer, due to our analysis method. Further evidence to this behavior is visible in the data, which do not show arbitrarily scattered data points but follow an undulating line, which is clearly visible for the agglomerate data at high pressures (pluses in the rightmost panel of Fig. \ref{fig:PPress}). For the other particle types, this behavior is only visible when the data points are plotted linearly.

Another issue of the type 1 ground experiments is that after the addition of another particle layer, the sample had to be moved to determine the new mass. Any associated vibrations may have caused restructuring and, thus, may have artificially increased the filling factor of the sample. The data of the mm-sized particles are again affected by the loss of particles, which stick to the walls at low and medium pressures, which increases the measured filling factor. Apart from this, our laboratory data fit well to our parabolic-flight results.

\section{Modeling}\label{sect:CE} 
\subsection{Regolith consisting of solid spherical particles}\label{sect:RCS}
\citet{Guettleretal2009} used the following empirical description for the pressure dependence of the filling factor of cohesive-particle layers:
\begin{eqnarray}
\Phi(\Sigma)=\Phi_2 -\frac{\Phi_2 - \Phi_1}{\exp\left(\frac{\log \Sigma -\log p_m}{\Delta} \right) +1}.\label{eq:gu}
\end{eqnarray}
Here, $\Sigma$ is the pressure and $\Phi_1$, $\Phi_2$, $p_m$, and $\Delta$ are the minimum and maximum filling factor at very low and very high pressures, the turnover pressure and the logarithmic width of the transition from low to high filling factor. The transition width was experimentally found by \citet{Guettleretal2009} to be $\Delta =0.58$ for omni-directional compression. %
%In an earlier paper, we showed that for spherical mono-disperse particles at very low pressures, when the regolith is in the hit-and-stick regime during its formation, the filling factor is $\Phi_1 = 0.15$ \citep{BS2004}. This value is also termed the filling factor of the random ballistic deposition (RBD).
The packing density $\Phi_1$ (Random Ballistic Deposition, RBD) has been measured by \citet{BS2004} which matches results from numerical simulations and corresponds to a coordination number of 2 for  spherical monodisperse particles, independent of their radius. RBD means that particles sequentially drop at random positions on a surface and stick where they hit. $\Phi_2$ (Random Close Packing, RCP) is used for spherical monodisperse particles independent of their size and corresponds to a mean coordination number of 6 (Yang et al. 2000). At RCP, a regolith is jammed and cannot be compressed further.

Sliding is not possible either, because sliding forces are very high and the regolith particles will crunch prior to sliding.
%Because the pressure in Regolith is a tensor where  the pressure has the directipon of the force applied to it with a slight deviation (friction angle), the particle contact force apart from the vertical direction is with a good approximation independent on the pressure inside the regolith and only depends on the particle contact force. Because the rolling force depends on the

At a filling factor of $\Phi = 0.15$, the average coordination number of the particles inside the regolith is $2$ \citep{vandelagemaatetal2001} so that the particles are free to move. Any pressure is transformed into a force at the particle contacts, which is proportional to the particle cross section and the inverse filling factor. With this consideration, the pressure at which the restructuring of the regolith starts, can be calculated by
\begin{eqnarray}
p_1 =  \frac{F_{roll} \Phi_1}{\pi~r_p^2}. \label{eq:pm}
\end{eqnarray}
with
\begin{eqnarray}
F_{roll}=F_0 \left(\frac{r_p}{r_0}\right)^{\frac 23} \label{eq:fro}
\end{eqnarray}
\citep{krijtetal2014}.  Here, $r_p$ is the particle radius.

Equation \ref{eq:pm} relates the pressure inside the regolith to the force exerted on a single particle inside the regolith that makes it roll over its contact points. To derive this force, we multiplied the ambient pressure inside the regolith by an effective cross section per unit monomer particle. This effective cross section (at a coordination number of 2) is the cross section of a monomer particle divided by the packing density, i.e. the smaller the packing density the higher the force on a single particle contact for a given pressure. In other words, $p_1$ is the maximal pressure a regolith can sustain without restructuring. The friction force
$F_{0}$ at the radius $r_0$ are taken from the measurement of \citet{Heim1999}. With the assumption that $\Delta$ is independent of particle size, the ratio
\begin{eqnarray}
R=\frac{p_m}{p_1}\label{eq:rpm}
\end{eqnarray}
is particle size independent and it is possible to calculate $p_{m}$
from $p_1$ for all particle sizes,
\begin{eqnarray}
p_m =  \frac{F_{roll} \Phi_1}{\pi~r_p^2} R.\label{eq:rpr}
\end{eqnarray}

The measurements by \cite{Guettleretal2009} show filling factors slightly below $\Phi =0.15$ at low pressures, which implies that their regolith at low pressures included void spaces. Therefore, it is not possible to use $p_1$ from Eq. \ref{eq:pm} in this case.
%Instead, we equate the pressure $p_1$ to the pressure at which in their Figure 2 a filling factor of $\Phi$=0.15 is reached.
%{\bf 1.12}
For $p_1$ we instead have to use the filling factor $\Phi =0.15$ to assure that the coordination number of the particles is 2. Therefore we take the pressure from the compression curve of \cite{Guettleretal2009} at $\Phi =0.15$ and associate it with $p_1$.  
A simple way to calculate $R$ is by inserting the corresponding values $p_m$ and $r_p$ from  a compression measurement (which is in our case the \cite{Guettleretal2009} omnidirectional compression from his Table 1 with the above corrections) in Equation \ref{eq:rpr2}.

We assume that the logarithmic transition width $\Delta$ is independent of particle size so that we obtain a constant value of $\Delta = 0.58$ and a constant R for all samples. In a descriptive way, this is the case, because the increasing resistivity against compression at increasing filling factors is due to a reordering of the particles and a corresponding increase of the mean coordination number. The mean coordination number is independent of the particles size  and is directly connected to the force chains  in a regolith, which are responsible for its resistance to compression. For a proof please compare figure 4 of \citet{Langemaat}, which was done for mono-disperse nanometer sized particles with Figure 9 of \citet{Yangetal2000} which is plotted for different sizes (1 $\mu$m to 1000 $\mu$m) depending on porosity. These curves are nearly identical.

\begin{figure}[!thb]
    \center
      \includegraphics[width=0.45\textwidth]{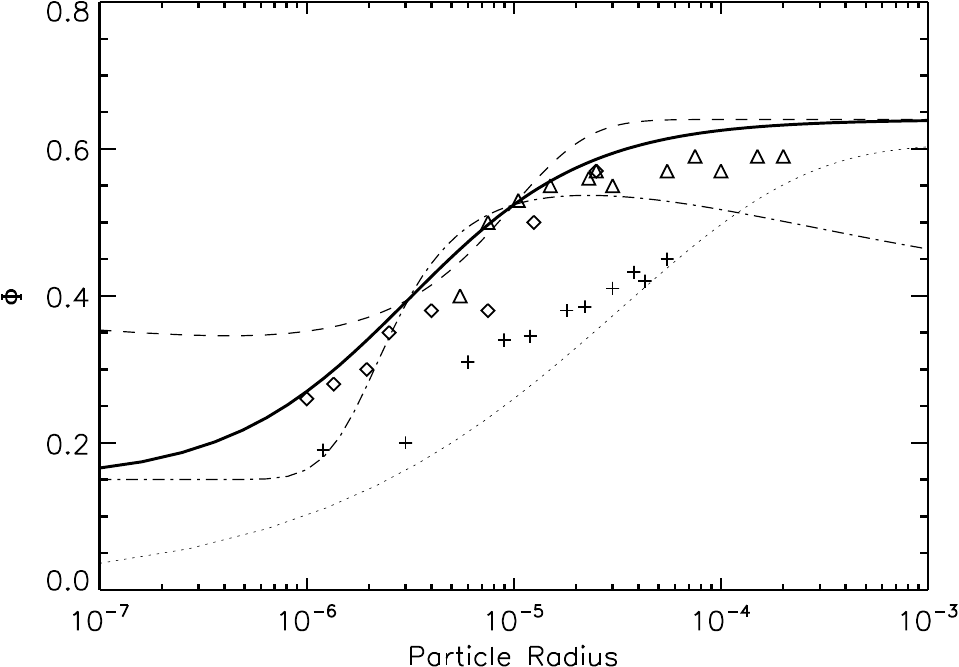}
    \caption{\label{fig:PoY} The dependency of the filling factor $\Phi$ on the particle radius at constant fill height and gravity. Pluses: data from irregular particles with a narrow size distribution \citep{Yu1997}. Diamonds: data from spherical mono-disperse particles \citep{Milewski}. Triangles: data from spherical mono-disperse particles \citep{Feng1998}. Solid curve: our model with constant $\Delta$. Dashed-dotted curve: our model with $\Delta \propto r^{0.5}$. Dashed line: our model with $\Delta \propto r^{-0.5}$.	     Dotted line: best fit of measurements with irregular particles of several authors taken from \citet{Yu2004}.}
\end{figure}

To prove the assumption of a constant $\Delta$ and to show the consistency of our model, we plotted our data for a constant ambient pressure $\Sigma$ (calculated from \citet{Yu1997} as described below) as a function of the particle radius (solid curve in Fig. \ref{fig:PoY}). This was done by evaluating the relation between $p_m$ and the particle radius through combining Eqs. \ref{eq:pm}, \ref{eq:fro} and \ref{eq:rpm},
%\begin{eqnarray}
%p_m(r_p)=F_0\Phi_1 r_0^{-\frac23} \pi^{-1} r_p^{-\frac43}. \label{eq:rpr2}
%\end{eqnarray}

\begin{eqnarray}
p_m(r_p)=F_0\Phi_1 r_0^{-\frac23} \pi^{-1} r_p^{-\frac43} R. \label{eq:rpr2}
\end{eqnarray}

The parameters of the resulting solid curve in Fig. \ref{fig:PoY} are shown in Table \ref{table:FFP}.

A potential radius dependency of $\Delta$ was realized by multiplying  $\Delta$ with
\begin{eqnarray}
\left(\frac{r}{r_j}\right)^{0.5}
\end{eqnarray}
or
\begin{eqnarray}
\left(\frac{r}{r_j}\right)^{-0.5} ,
\end{eqnarray}
with $r_j = 10$\ $\mu$m being a scaling parameter. The dotted amsnd dashed curves in Figure \ref{fig:PoY}) show the behavior of the filling factor as a function of particle radius for these cases. Both deviations from constant $\Delta$ result in a non-monotonic behavior of the curves, in contradiction to all available measurements and numerical simulations, which are monotonically increasing and do not fit these curves \citep[see, e.g. the data of several authors given in][and the data shown in Figure \ref{fig:PoY}]{Yu2004,Yangetal2000}.
The measurements by \citet{Yu1997} were performed with irregular particles with a narrow size distribution. Their data are shown as pluses in Fig. \ref{fig:PoY}). Their experiments and the corresponding setup are so well described that we could use their measurement results to calculate the pressure. \citet{Yu1997} used the standard method: they filled a truncated cone with an upper diameter of 57 mm, a lower diameter of 44 mm and a height of 70 mm to the rim with their particles and measured their weight. We calculated the mean pressure inside the container for the average mass they filled in for different particle sizes and therefore packing densities. We did this by integrating Equation \ref{eq:jan} over the fill height.

\citet{Feng1998} and \citet{Milewski} did measurements with spherical particles using the same method. The measurements of \citet{Yu1997} (diamonds in Fig. \ref{fig:PoY}) and \citet{Milewski} (triangles in Fig. \ref{fig:PoY}) fit our calculations properly, which shows that our Eq. \ref{eq:rpr2}, a constant $\Delta$, a power law for $p_m(r_p)$ as well as the ansatz with a Fermi function is appropriate. The measurements of  \citet{Yu1997} show that the power law also holds for irregular particles, and that irregular particles behave like spherical particles that are somewhat smaller. Quantitatively, this means that the corresponding $p_m$ of irregular grains is approximately a factor of 8 larger than for spherical particles so that the curves in Fig. \ref{fig:PoP} are shifted by a factor 5 to the right to be valid for irregular particles. With the above considerations, and having in mind that $p_m$ for regolith particles with a wide size distribution has the opposite effect on $p_m$ and nearly cancels the change from spherical to irregular grains (see the discussion in Section \ref{sect:LOOM}), this estimates a maximum error of our curves caused by the assumption of spherical particles.

The dotted curve in Fig. \ref{fig:PoY} is a best fit taken from \citet{Yu2004} to the measurements with irregular particles from several authors.

To derive $p_m$ from the measurements shown in Fig. \ref{fig:PPress}, we fit Eq. \ref{eq:gu} to the measured data, with $p_m$ being the only fit parameter. The resulting values for $p_m$ as well as the other parameters used in Eq. \ref{eq:gu} are shown in the first three lines of Table \ref{table:FFP}. The best-fitting curve is shown as a solid line in Fig. \ref{fig:PPress}. As the data show some considerable scatter, we also plot two envelope curves (shown as dashed lines in Fig. \ref{fig:PPress}), which define a maximally possible range for the transition pressure from $p_{\rm min}$ to $p_{\rm max}$. These two parameters are also given in Table \ref{table:FFP} and were derived such that they define the minimal range of transition pressures for which no data points fall outside the envelope curves. Due to the above-mentioned hysteresis effect and vibrations during handling in the laboratory experiments (see \ref{sect:LES}), which result in a too-low packing density for some of the data points and a compaction above RCP for others, respectively, the above conditions is strictly obeyed for the microgravity data, but we allow the laboratory data to fall outside the envelope curves.

\begin{table*}
        \centering
        \caption{\label{table:FFP}Parameters of the fit function (Eq. \ref{eq:gu}) for the four particle types used in this study (rows 1-4), extrapolations to mm-sized and cm-sized agglomerate layers (rows 5-6) as well as for other compression studies (rows 7-8). Please note that $\phi_1$ from the \citet{Machii2013} data was increased to fit to our measurements. Rows 9-11 show the parameters for the model curves with constant $\Delta$ (solid curve in Fig. \ref{fig:PoY}), with $\Delta \propto r^{0.5}$ (dotted curve in Fig. \ref{fig:PoY}) and with $\Delta \propto r^{-0.5}$ (dashed curve in Fig. \ref{fig:PoY}). All pressures are given in Pa.}
\begin{tabular}{l c c c c c c c c}
\hline
Particle type & $\phi_1$ & $\phi_2$ & $p_m$& $p_{min}$& $p_{max}$ & $\Delta$ & $\Sigma$ & Power\\
\hline
10~$\mu$m spheres  & 0.15 & 0.64 & 5$\times 10^3$ & 40 & 1.3$\times 10^5$ & 0.58  \\
60~$\mu$m spheres& 0.15 & 0.64 & 0.35& 0.02 & 16 & 0.58  \\
%60-$\mu$m ground & 0.15 & 0.64 & 2& 0.05 & 10 & 0.58  \\
1~mm spheres & 0.15 & 0.64 & 0.3& 0.005 & 10 & 0.58  \\
%1mm ground  & 0.15 & 0.64 & 0.15& 0.05 & 20 & 0.58  \\
100-250~$\mu$m agglomerates & 0.05 & 0.37 & 0.6 & 0.06 & 6 & 0.58  \\
%Agglomerates ground & 0.05 & 0.37 & 4 & 2 & 40 & 0.4  \\
1~mm agglomerates & 0.05 & 0.55 & $6.1\times10^{-2}$& & & 0.58\\
1~cm agglomerates & 0.05 & 0.55 & $4.7\times10^{-3}$ & & & 0.58\\
\citet{Guettleretal2009} & 0.12 & 0.58 &  $13 \times 10^3$ & & & 0.58\\
\citet{Machii2013} & 0.35 & 0.55 &  $4 \times 10^5$ & & & 0.58\\
Solid curve in Fig. \ref{fig:PoY} & 0.15 & 0.64 & & & & 0.58 & 357 Pa & 0 \\
Dotted curve in Fig. \ref{fig:PoY} & 0.15 & 0.64 & & & & 0.58 & 357 Pa & +0.5 \\
Dashed curve in Fig. \ref{fig:PoY} & 0.15 & 0.64 & & & & 0.58 & 357 Pa & -0.5 \\
\hline
\end{tabular}
\end{table*}

\begin{figure}[!thb]
    \center
      \includegraphics[width=0.45\textwidth]{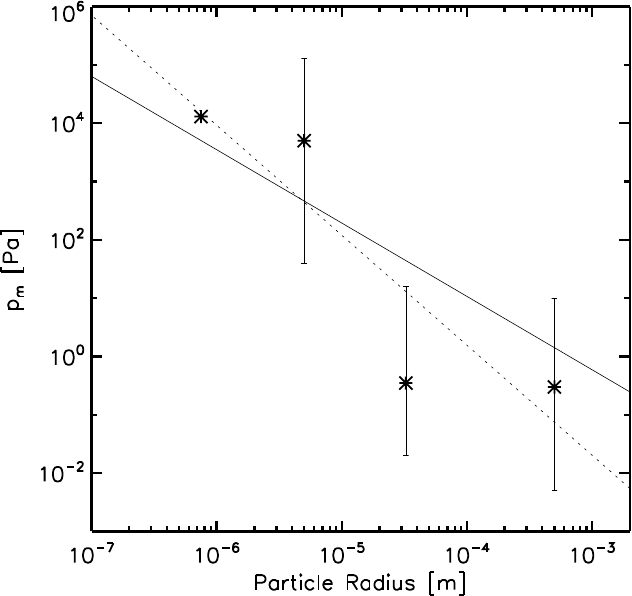}%\columnwidth
    \caption{\label{fig:PoR} The transition pressure p$_m$ from Eq. \ref{eq:gu} as a function of the radius of the spherical regolith particles. The solid line shows the model curve, given by Eq. \ref{eq:rpr2}. The data point for the smallest particle size stems from the compression measurements of  \citet{Guettleretal2009}. All other data points are derived by fitting Eq. \ref{eq:gu} to our measurements. The error bars denote the envelope curves in Fig. \ref{fig:PPress} and are given by  $p_min$ and $p_max$.}
\end{figure}

In Fig. \ref{fig:PoR}, all derived transition pressures $p_m$ and their respective uncertainty ranges, $p_{\rm min} - p_{\rm max}$ (all data identical to those of the first three rows in Table \ref{table:FFP}), are shown as asterisks with error bars as a function of the regolith-particle radius. The data point from the omnidirectional compression experiments of \citet{Guettleretal2009} is also shown in Fig. \ref{fig:PoR} at $r_p = 0.75 ~\rm \mu m$ (asterisk without error bar). The solid line in Fig. \ref{fig:PoR} corresponds to Eq. \ref{eq:rpr2}, i.e. $p_m \propto r_p^{-4/3}$. The dotted line is the best fit to the data assuming a power law and minimizing the chi-square error statistics of the logarithmic data points.

We think that this deviation in slope can be attributed to the rather large errors in our measurements, but is not proof that the model behavior (Eq. \ref{eq:rpr2}) is wrong.

The rough agreement between model (Eq. \ref{eq:pm}) and data indicates that rolling is the major effect at the particle level during compression and shows that it can describe our measurements reasonably well.
%The slight discrepancy for the 1~mm-sized particles can be understood in terms of the overcorrected particle loss to the walls, as explained above.
Although the $p_m$ was not used to fit the model to our data, the transition pressure $p_m$ measured by \citet{Guettleretal2009} for $\rm \mu m$-sized monomer grains is also reasonably well represented by the model.

\subsection{Regolith consisting of dust agglomerates}\label{sect:RFA}
At low pressures, the total filling factor of the regolith consisting of dust agglomerates should be
\begin{eqnarray}
\Phi_{total}=\Phi_{\mathrm{aggl}} \Phi_{\mathrm{RSS}},
\end{eqnarray}
with $\Phi_{\mathrm{\mathrm{aggl}}}$=0.35 \citep{Weidling2012} and $\Phi_{RSS}$ being the internal filling factor of the 100-250~$\mu$m-sized dust agglomerates and of the regolith super structure, respectively. The force on a single dust agglomerate inside the regolith can be calculated from the pressure $\Sigma$ within the regolith, analogous to Eq. \ref{eq:pm}, by
\begin{eqnarray}
F_{\mathrm{\mathrm{aggl}}}=\frac{\Sigma ~ \pi ~ r_{\mathrm{aggl}}^2}{\Phi_{RSS}}.
\end{eqnarray}
We assume that the friction against rolling-force for an agglomerate compared to solid particles of the same size is smaller by a factor of $\Phi_{\mathrm{aggl}}$ and the regolith super structure was formed by random ballistic deposition, which yields a filling factor of $\Phi_{RSS}$=0.15. We can then calculate the minimum pressure at which rolling occurs
\begin{eqnarray}
p_{roll}=\frac{F_{\mathrm roll} ~ \Phi_{\mathrm RBD}}{\pi ~ r_{\mathrm{aggl}}^2 ~ \Phi_{\mathrm{aggl}}}.
\end{eqnarray}
We get $p_{\mathrm roll} \approx 10^{-2}~{\mathrm{Pa}}$ and this is the pressure at which restructuring of regolith consisting of agglomerates should start. This is consistent with our measurements shown in Fig. \ref{fig:PPress} (right panel) and the analytical approximation shown in Fig. \ref{fig:PPress} (solid lines).

\subsection{Limits of our Model}\label{sect:LOOM}
Our model is based on measurements on regolith for spherical grains with radii between $r_p = 7 \times 10^{-7}$ m and $r_p = 5 \times 10 ^{-3}$ m and pressures in between $\Sigma = 10^{-7}$ Pa and $\Sigma = 10^{6}$ Pa. We apply our model to celestial objects for which the representative regolith-grain sizes modeled to be between $r_p = 4.4 \times 10^{-5}$ m and $r_p = 4.2 \times 10 ^{-2}$ m \citep{Gundlachetal2011}. The pressures at the second regolith-particle layer in these objects are between $\Sigma = 10^{-2}$ Pa and $\Sigma = 10$ Pa and can reach values of $\Sigma = 10^{3}$ Pa deep inside the regolith. Thus, the particle sizes have to be extrapolated by only one order of magnitude using the results of our experiments and the pressures are well within our measurement range. The fact that we only measured at a minimum $g$-level of 0.06 ms$^{-2}$ and, e.g., the asteroid 1996FG3 possesses a $g$-level of 3.3$\times$10$^{-4}$ ms$^{-2}$ at its surface is not a severe restriction of applicability, because only the pressure inside the regolith is relevant.

However, regolith particles on planetary objects are not mono-disperse spherical grains but are irregular in shape and possess some size distribution, as can be seen in the Apollo samples \citep{LunarSourcebook}. To test the influence of grain irregularity and poly-dispersity on the mechanical properties of the regolith, we previously already did unidirectional compression measurements with monodisperse and polydisperse irregular grains \citep{BSDR2006} with a size range of 0.1$\mu$m - 10$\mu$m \citep[see][their Figure 3]{Kothe2013}.
We found that the filling factor for RBD drops from $\Phi_1 = 0.15$ for monodisperse spherical grains to $\Phi_1 = 0.11$ and $\Phi_1 = 0.07$ if mono-disperse irregular and poly-disperse irregular grains are used \citep{BSDR2006}. The maximum filling factor at high unidirectional compressions  are $\Phi_2 = 0.33$ for mono-disperse spherical and irregular grains and drop to $\Phi_2 = 0.20$ for poly-disperse irregular particles \citep{BSDR2006}. Mind that these values are much lower than the RCP value, because under unidirectional compression the samples creep sideways. Although not explicitly given in our previous paper, the values for $p_m$ are almost identical for the three types of grains and read $p_m = 5 \times 10^3$ Pa, $p_m = 2 \times 10^3$ Pa, and $p_m = 4 \times 10^3$ Pa for the monodisperse spherical, the mono-disperse irregular and the poly-disperse irregular grains, respectively.

The measurements of \citet{Yu1997} show that irregular particles behave like spherical grains that are a factor of 5 smaller (see Section \ref{sect:RCS}). Applying Eq. \ref{eq:rpr2}, we find a corresponding change of $p_m$ by a factor of 8. However, as a size distribution \citep[as e.g. found by][]{Miy2007} reduces this effect on $p_m$ by a factor of 2 (see above), we therefore conclude that the morphology and poly-dispersity of the grains play only a minor role for the compression behavior of regolith.

The stratification measurements of the densities of lunar regolith show external packing densities larger than RCP \citep{LunarSourcebook}. This indicates that polydisperse grains can be compacted beyond this limit and show that our model underestimates the packing densities by approximately 12\% (discussed later).

Regolith from particles larger than 100 $\mu$m is sensitive to vibrations due to the large inertia of the particles compared to their inter-particle van der Waals forces. Thus, vibrations caused by impacts on the celestial objects can increase their regolith packing density by 0.1 \citep[][their Figure 5]{Yu1997}.

We do not think that sintering is an important issue for regolith particles, because these grains were produced by impact destruction and size selective re-accretion. Whether the re-accreted particles are directly formed by impact fragmentation or by fragmentation and re-sintering, is not of relevance. The latter is more realistic as the investigations in the Lunar Sourcebook \citep{LunarSourcebook} show that the regolith particles have a internal packing density of about 79\% from void internal spaces. During the formation of comets in the early solar nebula, the temperatures were presumably too small for relevant sintering processes, so that sintering will not have any influence for the stratification of cometary surface regolith. However, laboratory measurements of \citet{Pat-El2009} show that sintering during an orbital passage of the comet can increase the stability of the regolith in the uppermost few centimeters. This means that our model may not be applicable to the uppermost centimeters below the comet surface. However, as during each orbit a comet can lose surface material of a few meter depth, it is not clear whether heat-flow induced sintering is of relevance.

\section{Applications to Planetary Sciences}\label{sect:appli}
\subsection{Rocky objects}\label{sect:Rocky}
In this Section, we will derive the filling factor of the regolith on atmosphere-free planetary bodies as a function of depth below the surface. With this result, we will calculate the mechanical yield of the regolith under a mass of 100~kg resting on the surface on feet with a total area of 0.07~m$^2$. We will consider the effect of ambient gravity only and will disregard the compaction of the regolith by high-velocity impacts. In that course, we will compare our results with measurements taken by the Apollo 15-17 core tubes and drill cores on the Moon.

We again start with the calculation of the pressure inside the regolith of an arbitrary planetary body as a function of depth $h$ below the surface, $\Sigma(h)$.
Because the function $\Sigma(h)$ is non algebraic, we obtain its inverse
$h(\Sigma)$ by integrating the ansatz

\begin{eqnarray}
\dif h = \frac{\dif \Sigma(h)}{\rho_b g \Phi(\Sigma)},
\end{eqnarray}
using $\Phi(\Sigma)$ from Eq. \ref{eq:gu}, and get

%\begin{eqnarray}
%&&h=\frac{\Sigma}{g \rho_B (\Phi_1\Phi_2-2\Phi_2^2)} \Big\{\Phi_1-2 \Phi_2+ (\Phi_2-\Phi_1)\times \\ \nonumber
%&&\mbox{H2F1}\left[\Delta \ln(10),1,1+\Delta \ln(10),\frac{\Phi_2 \Sigma^{\frac{1}{\Delta\ln(10)}}
%p_m^{-\frac{1}{\Delta \ln(10)}}}{\Phi_1-2\Phi_2}\right]\Big\}, 
%\end{eqnarray}

\begin{eqnarray}
&&h=\frac{\Sigma}{g \rho_B (\Phi_1\Phi_2-2\Phi_2^2)} \Big\{\Phi_1-2 \Phi_2+ (\Phi_2-\Phi_1)\times \\ \nonumber
&&\mbox{H2F1}\left[\Delta \ln(10),1,1+\Delta \ln(10),\frac{\Phi_2 \Sigma^{\frac{1}{\Delta\ln(10)}}
p_m^{-\frac{1}{\Delta \ln(10)}}}{\Phi_1-2\Phi_2}\right]\Big\},
\end{eqnarray}

with H2F1 being the Gaussian hypergeometric function. Inserting the inverse function of Eq. \ref{eq:gu},

\begin{eqnarray}
\Sigma(\Phi)=p_m\left(\frac{\Phi_2-\Phi_1}{\Phi_2-\Phi}-1\right)^{\Delta \ln(10)}, 
\end{eqnarray}

we get

%\begin{eqnarray}
%&&h=\frac{p_m}{g \rho_B(\Phi_1\Phi_2-2\Phi_2^2)} \left(\frac{\Phi_1-\Phi}{\Phi-\Phi_2}\right)^{\Delta \ln(10)}\times\\ \nonumber
%&&\Big\{\Phi_1-2\Phi_2+(\Phi_2-\Phi_1)\times \\
%&&\mbox{H2F1}\left[1,\Delta \ln(10),1+\Delta \ln(10),\frac{ \Phi_2(\Phi_1-\Phi)}{(\Phi_1-2\Phi_2 )(\Phi-\Phi_2)}
% \right] \Big\} \label{eq:h}  \nonumber. 
%\end{eqnarray}

\begin{eqnarray}
&&h=\frac{p_m}{g \rho_B(\Phi_1\Phi_2-2\Phi_2^2)} \left(\frac{\Phi_1-\Phi}{\Phi-\Phi_2}\right)^{\Delta \ln(10)}\times\\ \nonumber
&&\Big\{\Phi_1-2\Phi_2+(\Phi_2-\Phi_1)\times \\
&&\mbox{H2F1}\left[1,\Delta \ln(10),1+\Delta \ln(10),\frac{ \Phi_2(\Phi_1-\Phi)}{(\Phi_1-2\Phi_2 )(\Phi-\Phi_2)}
 \right] \Big\} \label{eq:h}  \nonumber.
\end{eqnarray}

\begin{figure}[!tb]
    \center
      \includegraphics[width=.45\textwidth]{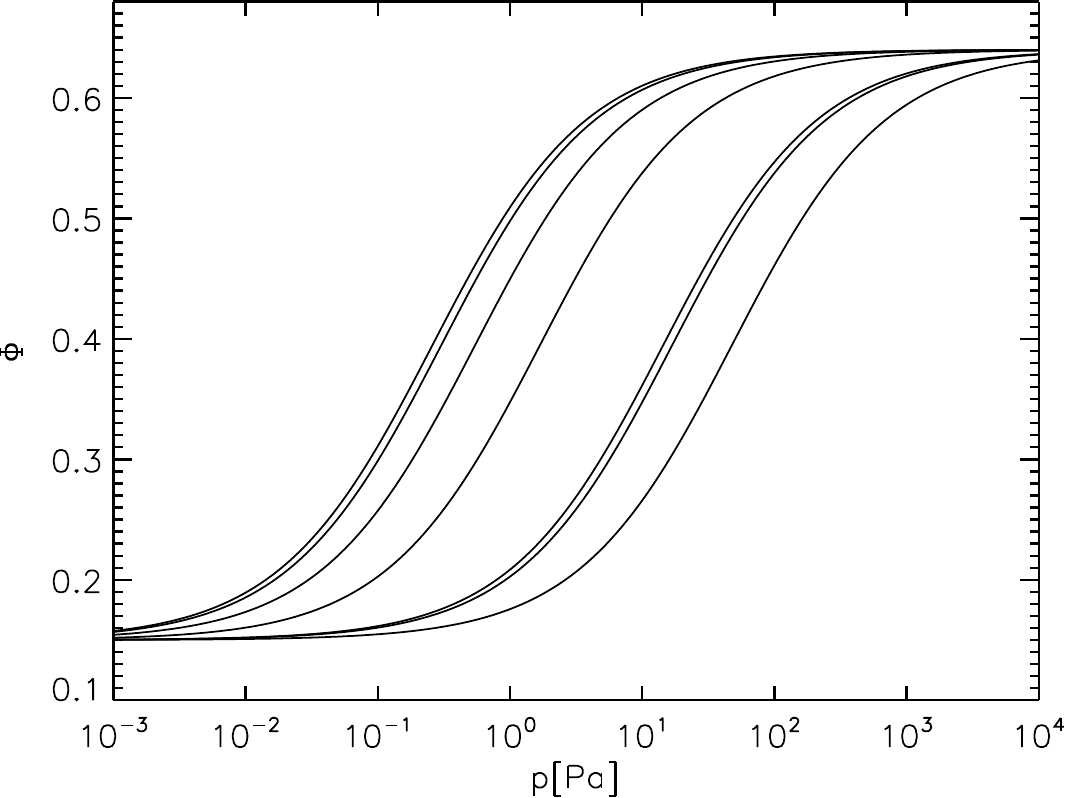}%\columnwidth
    \caption{\label{fig:PoP}The filling factor of the regolith of various Solar-System bodies as a function of hydrostatic pressure. The curves from  left to right Phobos (2.2~mm), asteroid 1996FG3,  asteroid Steins, asteroid Dodona, asteroid Vesta, the Moon and planet Mercury respectively.}
\end{figure}

\begin{figure}[!thb]
    \center
      \includegraphics[width=.43\textwidth]{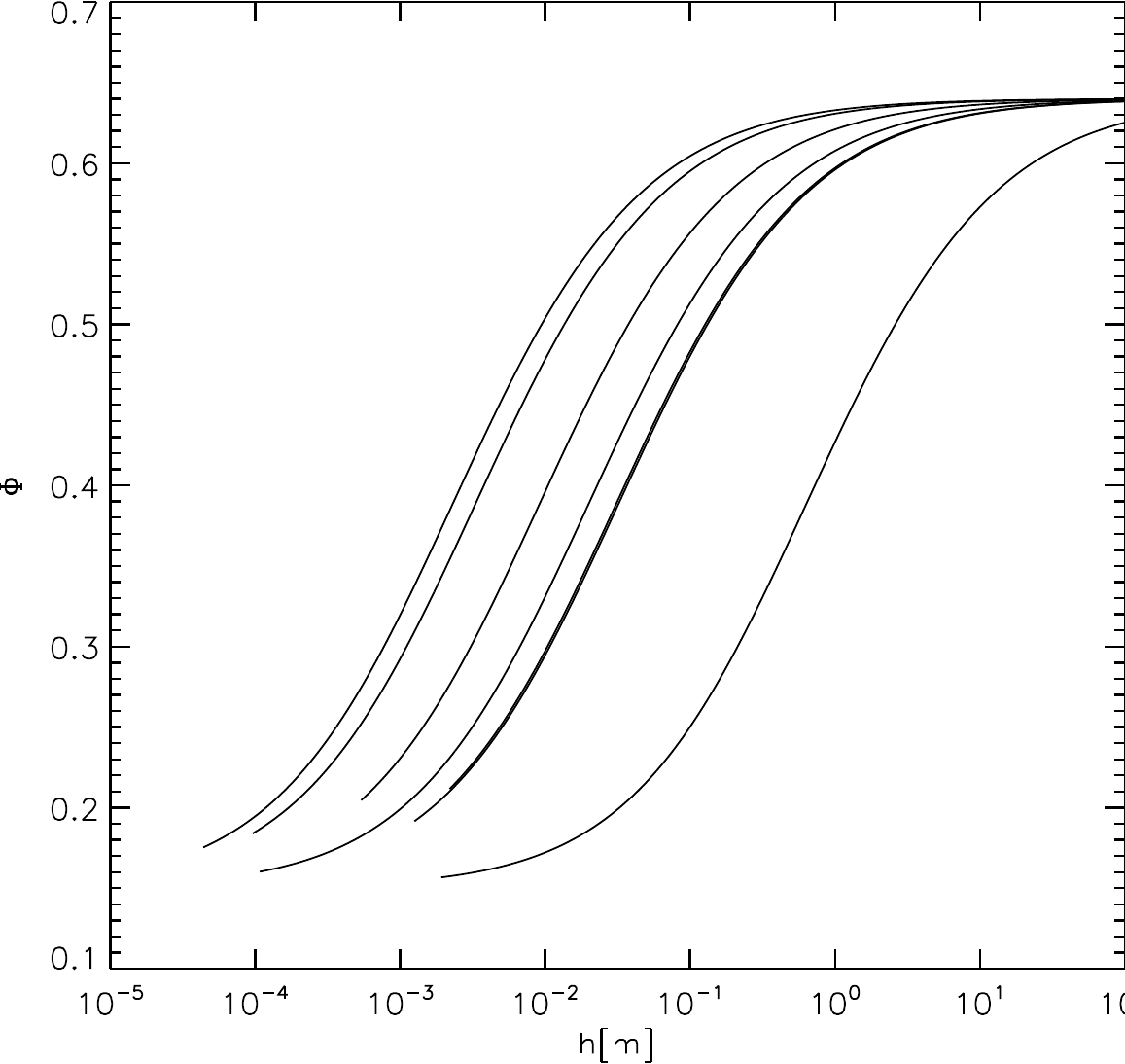}%\columnwidth
    \caption{\label{fig:HOP} The filling factor as a function of depth of the regolith (with particle diameters given in parentheses) of (from top to bottom) Mercury (44~$\mu$m), the Moon (96~$\mu$m), Dodona (0.6~mm), Vesta (108~$\mu$m), Phobos (2.2~mm), Steins (1.3~mm), 1996FG3 (2.0~mm), respectively.}
   %     the Moon (96~$\mu$m), Mercury (44~$\mu$m), Dodona (0.6~mm), Itokawa (4.2~cm), Vesta (108~$\mu$m), Phobos (2.2~mm), Steins (1.3~mm), and 1996FG3 (2.0~mm), respectively. Where the curves of Steins and Phobos are almost identical but start at different depth.}
  %   Vesta (108~$\mu$m), the Moon (96~$\mu$m), Mercury (44~$\mu$m), Dodona (0.6~mm), Itokawa (4.2~cm), Phobos (2.2~mm), Steins (1.3~mm), and 1996FG3 (2.0~mm), respectively. The curves start from a layer depth of one particle diameter. Regolith-particle sizes were taken from \citet{Gundlach2012}.}
\end{figure}

The filling factor as a function of pressure and depth below the surface of the regolith is shown in Figs. \ref{fig:PoP} and \ref{fig:HOP}, respectively, for a variety of planetary bodies with different gravitational accelerations $g$. The particle sizes of the regolith were taken from \citet{Gundlach2012}. For the graphs shown in both Figures, we used $\Phi_1 = 0.15$ and $\Phi_2 = 0.64$. $p_{\rm m}$ is calculated from the particle diameters of the regolith, using Eqs. \ref{eq:pm} and \ref{eq:rpm} (see Fig. \ref{fig:DFF}). The graph in Fig. \ref{fig:HOP} is the inverse of Eq. \ref{eq:h}.

\begin{figure}[!thb]
    \center
      \includegraphics[width=.45\textwidth]{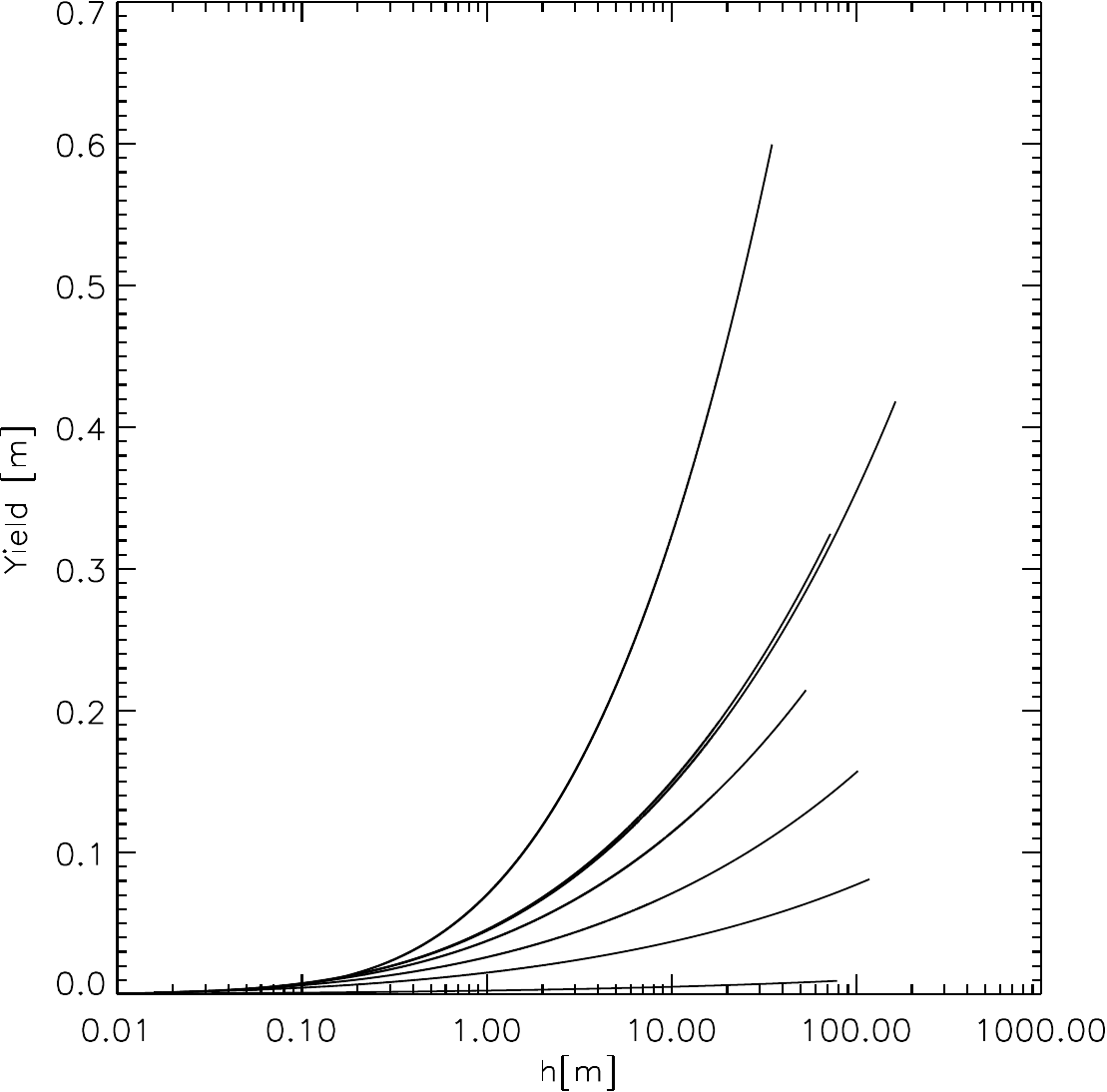}%\columnwidth
    \caption{\label{fig:HY} The yield of an object with a mass of 100~kg resting on feet with a total area of 0.07~m$^2$ (astronaut on one foot, small spacecraft) on different celestial bodies. The graphs give the yields (left to right) on
    1996FG3, Phobos, Steins (almost identical to Phobos), Vesta, Dodona, the Moon and Mercury,  respectively, as a function of the depth of the regolith layer. Please note that due to the friction angle, the yield will not increase further at a regolith depth at which the pressure is reduced to the elastic limit. The curves terminate at this point.}
\end{figure}

Because all reasonable loads, like, e.g. a 100~kg object on the surface with a foot-area of 0.07~m$^2$, lead to pressures that are orders of magnitude larger than the pressures needed to compact the regolith to RCP, we did the ansatz for the total yield of the regolith
\begin{eqnarray}
h_{\mathrm{yield}}=\int_0^{H_{\infty}}   \dif H  - \frac1{\Phi_{RCP}} \int_0^{H_{\infty}}  \Phi(H)  \dif H . \label{eq:yield}
\end{eqnarray}
Because Eq. \ref{eq:h} is monotonically increasing and continuous, we achieve the primitive of the inverse function of Eq. \ref{eq:yield}
(second integral) with the substitution $H=h(\Phi)$ and partial integration, i.e.
\begin{eqnarray}
\int_0^{H_{\infty}}\Phi(H) \dif H&={H_{\infty}} \Phi(H_{\infty})-0 \Phi(0) \label{eq:pip} \\
&-\int_{\Phi_0}^{\Phi_{\infty}} h(\Phi) \dif \Phi.\nonumber
\end{eqnarray}
Executing the integration of the first term in Eq. \ref{eq:yield} and inserting Eq. \ref{eq:pip} results in
\begin{eqnarray}
h_{\mathrm{yield}}=\int_{\Phi_0}^{\Phi_{\infty}}h(\Phi) \dif \Phi,\label{eq:yieldf}
\end{eqnarray}
which we solved numerically. The above term converges very slowly with $\Phi_{\infty} \rightarrow \Phi_{RCP} = 0.64$. This implies that $h_{\mathrm{yield}}$ depends on the regolith layer depth above a solid planetary interior. However, granular regolith has the characteristic to distribute point loads on its surface in a cone-shaped manner to a larger and larger area beneath. These cones have typically a friction angle between 35 and 50 degrees \citep{GranularDynamics}. This means that after a given depth inside a granular media, the force is being distributed over a large area. If the corresponding pressure reaches the elastic limit of the regolith, the compaction terminates. Here, we assume that the friction angle is 35 degrees and that the elastic limit of the regolith is given by Eq. \ref{eq:pm}. In Fig. \ref{fig:HY}, we show the yield as a function of the layer depth (using Eq. \ref{eq:yieldf}) for different planetary bodies. We plot the curves only for depths for which the pressure is above the elastic limit of the regolith. As the depth of the regolith is basically unknown, the maximum possible yield is given by the right end of each respective curve. However, if the regolith layer is shallower than that and is followed by bare rock beneath, the respective yields are smaller and can be derived from the curves in Fig. \ref{fig:HY}. One can recognize from Fig. \ref{fig:HY} that even for the deepest possible regolith depths, the static yield of the regolith under the weight of the assumed spacecraft or astronaut is  on the order of a few decimeters.

To compare our results with measurements taken on the Moon with Apollo 15-17 core tubes given in Table 9.5 of \citet{LunarSourcebook}, we average our lunar data of Fig. \ref{fig:HOP} over the depth ranges given in \citet{LunarSourcebook}. The porosity values of \citet{LunarSourcebook} include the internal porosity of the lunar regolith grains. This internal packing density was measured and reads $\sim$0.79   \citep{LunarSourcebook}. In Table \ref{table:PDM} we present our data on the filling factor of the lunar regolith together with the external packing density, corrected for the internal packing density of the grains \citep[both taken from ][]{LunarSourcebook}. The lunar surface regolith obviously is packed sightly denser than predicted by our model. This is because of the size distribution and irregularity of the lunar regolith particles for which packing densities up to 70\% are possible \citep{Desmond2013}.

\begin{table}
        \centering
        \caption{\label{table:PDM}Comparison of our predicted filling factor (center column) of the lunar regolith (averaged over the depth range shown in the left column) with measured external filling factors from Apollo published in \citet{LunarSourcebook} (right column).}
\begin{tabular}{l c c c c c c}
\hline
Depth Range (m) & Our Model & Measured \\
\hline
0 - 0.15 & 0.57 & 0.60  \\
0 - 0.30 & 0.59 & 0.64  \\
0.30 - 0.60 & 0.62 & 0.70 \\
0 - 0.60 & 0.60 & 0.68  \\
\hline
\end{tabular}
\end{table}

\citet{Magri2001} estimated the porosity of the first meter of regolith on asteroids using ground-based radar measurements. However, their data possess huge errors and it is not possible to find a correlation between filling factor and size of the asteroids in their data. If we average over all their findings, we get a filling factor of 0.49 $\pm$ 0.14. If we assume that the regolith grains of the asteroids possess the same internal packing density as measured for the lunar regolith particles, the average packing density is 0.62 $\pm$ 0.14, which matches with our model that predicts that at a few decimeters depth the regolith is at RCP. This estimations are, however, not a strong confirmation of our model, because the averaging over objects of different sizes and taxonomic types might not be appropriate.

\subsection{Comet 67P/Churyumov-Gerasimenko}
According to the model of \citet{Skorov2012}, comet nuclei consist of dust and ice agglomerates, each consisting of $\mu$m-sized solid dust and ice grains. To calculate for this model the pressure, depth and yield-load dependencies of the filling factor on a comet, we follow the calculation in Section \ref{sect:Rocky}. Here, we assume that ice agglomerates have a transition pressure that is a factor of 10 larger than that of our silicate agglomerates \citep{Gundlachetal2011}. We calculated the comet-regolith compression for agglomerates of 1~cm, 1~mm, 180~$\rm \mu m$ in size, respectively, the latter being the dust-aggregate size in our experiments. %For this, we assumed that the transition pressure for agglomerates has the same radius dependency as for solid grains because

Particles that are in contact possess a contact surface whose size is proportional to their radius squared. During rolling, additional contact surface is added in the direction of rolling and removed in the opposite direction. However, this process of creation and removal of contact surface is asymmetric, because the contact forces pull the contacting surfaces outward before they break. This difference causes a torque on the particle, which is known as rolling friction \citep[see][]{krijtetal2014,DoTi1995}. The radius dependence of this mechanism is found by \citet{krijtetal2014} to be given by r$^{\frac23}$. Agglomerates in contact possess a contact area for which the number of monomer particles is proportional to the agglomerate radius squared. During rolling, new particle-particle contacts are formed in the direction of rolling and broken in the opposite direction, because the monomer particles are also pulled outward. Because of this analogy between solid-solid and agglomerate-agglomerate contacts, we assume that the agglomerate-radius dependence of this mechanism follows the same power law as for compact particles. The parameters used in our calculations are also given in Table \ref{table:FFP}.

In the following, we apply our calculations to comet 67P/Churyumov-Gerasimenko, which is being visited by the Rosetta spacecraft in 2014/2015. As the Rosetta-mission data concerning the g-levels have not been published at the submission date of this paper, we calculated them using the published mass, average mass density and volume of comet 67P/Churyumov-Gerasimenko \citep{Sierksetal2015} of $m = 10^{13}$~kg, $\rho = 470 \rm ~kg~m^{-3}$ and $V = 25 ~ \rm km^3$. Thus, we get for the average radius of $R = 1.8$~km a surface acceleration of $g = 2 \times 10^{-4} \rm ~m~s^{-2}$.

\begin{figure}[!thb]
    \center
     \includegraphics[width=.47\textwidth]{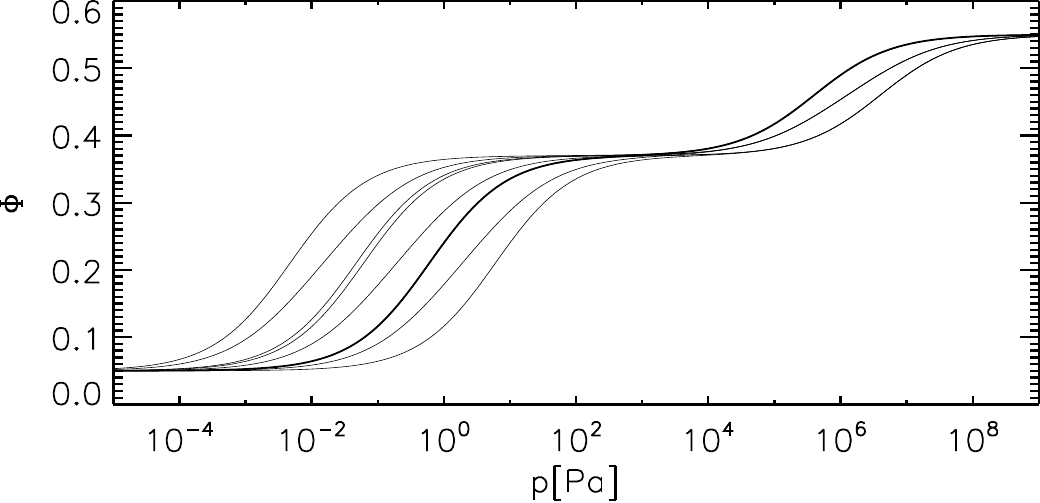}%\columnwidth
    \caption{\label{fig:PhiH} Filling factor as a function of pressure of cometary material. The first group of three curves represents regolith from cm-sized agglomerates, the second group of three curves represents regolith from mm-sized agglomerates, and the third group of three curves represents regolith from 180~$\mu$m-sized agglomerates, respectively. The high-pressure behavior of the material is based on data from \citet{Machii2013}. These curves are  independent of the agglomerate size of the regolith. Within these groups, the
    left curve is regolith from dusty material, the middle curve is a 50\%-50\% (in volume) mixture of agglomerates from dust and water ice, and the right curve is regolith from water ice. Please note that the curve of the ice particles of the mm-sized agglomerates is almost identical with the curve of the dust particles of the 180 $\mu$m sized particles.}
\end{figure}

We calculated the properties of cometary material for two extreme cases, namely that he comet consists of dust agglomerates or ice agglomerates only, and one average case that the comet consists of a 50\%-50\% (in volume) mixture of dust and ice agglomerates, as proposed by \citet{Skorov2012}. We further assume that the dust and ice agglomerates are arranged in random or alternate order so that the compressive behavior can be averaged in the following way
\begin{eqnarray}
\Phi_{\mathrm{ice+dust}}(p)&=&\frac12 \left(\Phi_{\mathrm{ice}}(p)+\Phi_{\mathrm{dust}}(p)\right) \\
\mathrm{Yield}_{\mathrm{ice+dust}}(p)&=&\frac12 \left(\mathrm{Yield}_{\mathrm{ice}}(p)+\mathrm{Yield}_{\mathrm{dust}}(p)\right).
\end{eqnarray}
Fig. \ref{fig:PhiH} shows the filling factor as a function of pressure for comet-nucleus material, assuming that the comet consists of agglomerates of 1~cm size (first group of 3 curves), 1~mm size (second group of 3 curves) and 180~$\mu$m size (third group of 3 curves). Within these groups, the agglomerates consisting of dust agglomerates are always represented by the left curve, the ice-and-dust mixture by the center curve, and the ice agglomerates by the right curve, respectively.

Regolith consisting of agglomerates has a smaller total RBD filling factor compared to regolith from solid grains, due to the intrinsic porosity of the agglomerates (see Section \ref{sect:RFA}). Therefore, they can be compacted further than RCP of the regolith super structure by destruction or deformation of the spherical agglomerates and subsequent compaction to a global RCP structure. Its strength against compression is higher than that of a regolith consisting of compact grains of the same size \citep{Machii2013}. The compression curve thus consists of two parts. The first compaction stage is due to the reorientation of the agglomerates by rolling until RCP of the super structure is reached (our measurements in Section \ref{sect:Expres}). The data for the second compaction stage are based on the compression measurement of \citet{Machii2013}, which follow the compression of the agglomerates themselves. This compression takes places at pressures larger than $10^5$ Pa, because the $\mu$m-sized grains (i.e. the constituent grains of the agglomerates) determine the compression strength. The original data of \citet{Machii2013} start at a filling factor of 0.24, whereas the maximum compression of our measurements ends at a filling factor of 0.37. This is due to the different preparation methods of the agglomerates. The more spherical and monodisperse the agglomerates are, the smaller is their filling factor at RCP \citep{Donev2004}. We used the results by \citet{Machii2013} by increasing their initial filling factor to our end value of 0.37 (see Table \ref{table:FFP}) and assumimg that the pressure required for the compression of icy agglomerates to a given filling factor is 10 times higher than for dusty ones. For compressions where the agglomerates are in RCP, the total filling factor as a function of pressure is independent of the agglomerates size. This means that the curve corresponding to the \citet{Machii2013} data (slopes at high pressures) is valid for all three agglomerate sizes. An explanation for this behavior is given in the Appendix.

Mind that the pressures at the second compression stage are much higher than the internal pressures on small Solar-System objects. Such pressures can occur on a comet only during high-velocity impacts. For impact pressures in the GPa range we recommend the work of \cite{Beitz2013}.

The high rolling and sticking forces of water ice lead to a comet nucleus of
low filling factor; whereas the regions close to the comet surface presumably consist of dust agglomerates only, due to the sublimation of water ice, and will thus possess a higher filling factor. A mixture of dust and ice agglomerates possibly results in a filling factor in between the extreme values (see Fig. \ref{fig:PhiH}). A comet nucleus is assumed to possess a few centimeter thick layer of pure dust agglomerates, which is followed by a mixture of dust and ice agglomerates. Thus, the filling factor will follow the respective left curve of one of the three groups in Fig. (\ref{fig:PhiH}) in the upper centimeters of the surface, followed by the center curve for a 50\%-50\% (in volume) mixture of dust and ice. For a higher ice content, the filling factor will be closer to the pure ice curve, which is the right-most curve in the respective triples. For a lower ice abundance, the filling factor will be closer to the left curve. The three triple-curves in Fig. \ref{fig:PhiH} represent dust and ice agglomerate sizes of 1~cm, 1~mm, and 180~$\rm \mu m$, respectively.

\begin{figure}[!thb]
    \center
         \includegraphics[width=.45\textwidth]{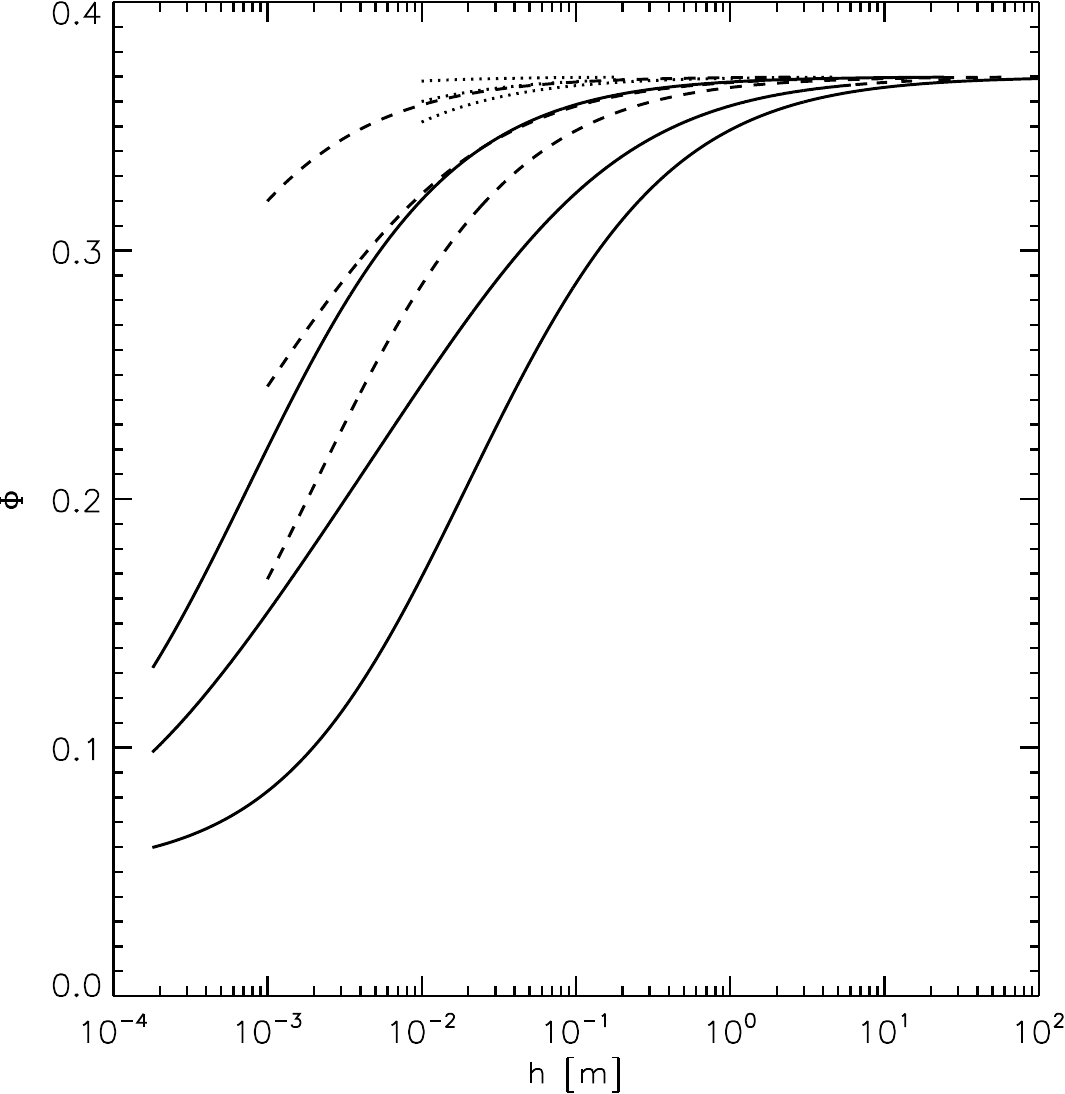}%\columnwidth
    \caption{\label{fig:POHC} The filling factor as a function of depth below the
    surface of Comet 67P/Churyumov-Gerasimenko. The  group of three dotted curves represents particle layers consisting of  1~cm-sized agglomerates, the  group of dashed graphs represents particle layers consisting of 1~mm-sized agglomerates,  the  group of solid graphs represents particle layers consisting of 180~$\mu$m-sized agglomerates. Within these groups, the upper curve represents dusty material, the middle curve a 50\%-50\% (in volume) mixture of agglomerates from dust and water ice, and the lower curve  ice agglomerates, respectively. All curves start from a depth of one agglomerate diameter.}
\end{figure}

Applied to comet 67P/Churyumov-Gersimenko, we get a filling factor as a function of depth as shown in Fig. \ref{fig:POHC}. The respective triple curves are, those for agglomerate sizes of 1~cm (dotted lines), 1~mm (dashed lines), and 180~$\rm \mu m$ solid lines. The respective upper curve represents agglomerates consisting of dust, the center curve represents 50\%-50\% (in volume) mixtures of dust and ice, and the lower curve stands for pure ice agglomerates, respectively. All curves start from a layer depth of one particle diameter.

To apply our results to the anticipated landing of the Rosetta lander on comet 67P/Churyumov-Gerasimenko, we calculated how much the surface of the comet nucleus will yield at touchdown. For the lander, we assume a mass of 100~kg and a total landing-feet area of 0.07~m$^2$. In case the spacecraft rests on the surface it will only render the agglomerate super structure to RCP and will not considerably sink in under its own weight, because the pressure below its feet is only $\sim0.4$~Pa and, thus, much lower than the pressure needed to compact the regolith (see Fig. \ref{fig:PhiH}). However, the impact pressure will be higher than the static weight of the spacecraft.

Instead of using an impact model \citep[e.g.][]{Melosh1998}, we estimate the impact pressure by assuming that the impact velocity will be decelerated steadily by the landing gear on a length of 10~cm, which is foreseen for the Rosetta lander. This dynamic pressure is treated as a static pressure on the surface. According to \citet{Melosh1998}, dynamic compression of our regolith would require a 20 times stronger deceleration and does therefore not appear. An issue of the static ansatz is the small deceleration time (0.4 s), because of the low sound speed in the regolith of about 20 ms$^{-1}$ \citep[see][]{Beitz2013}. This limits the depth to which the regolith can be compressed to 8 m. But due to the friction angle, the elastic limit is already reached for 180~$\mu$m-sized  aggregates at 3 m, 7 m and 10 m for pure ice, dust-ice mixture and pure dust, respectively, assuming an impact velocity of 1~m~s$^{-1}$. %For the 180 $\mu$m agglomerates this depth limitation is taken into account in the yield-calculations  below.
For the mm and cm-sized agglomerates the elastic limit is reached at a depth larger than 100~m by the gravitational acceleration of the comet only. For these agglomerate sizes the yield is independent of the impact velocity.

%In case the comet consists of mm and cm-sized agglomerates the penetration depth is trivial and below a few cm for all reasonable impact velocities, because the regolith on the surface is nearly at RCP for dust and ice agglomerates in this case (see Fig. \ref{fig:POHC}).

\begin{figure}[!thb]
    \center
         \includegraphics[width=.45\textwidth]{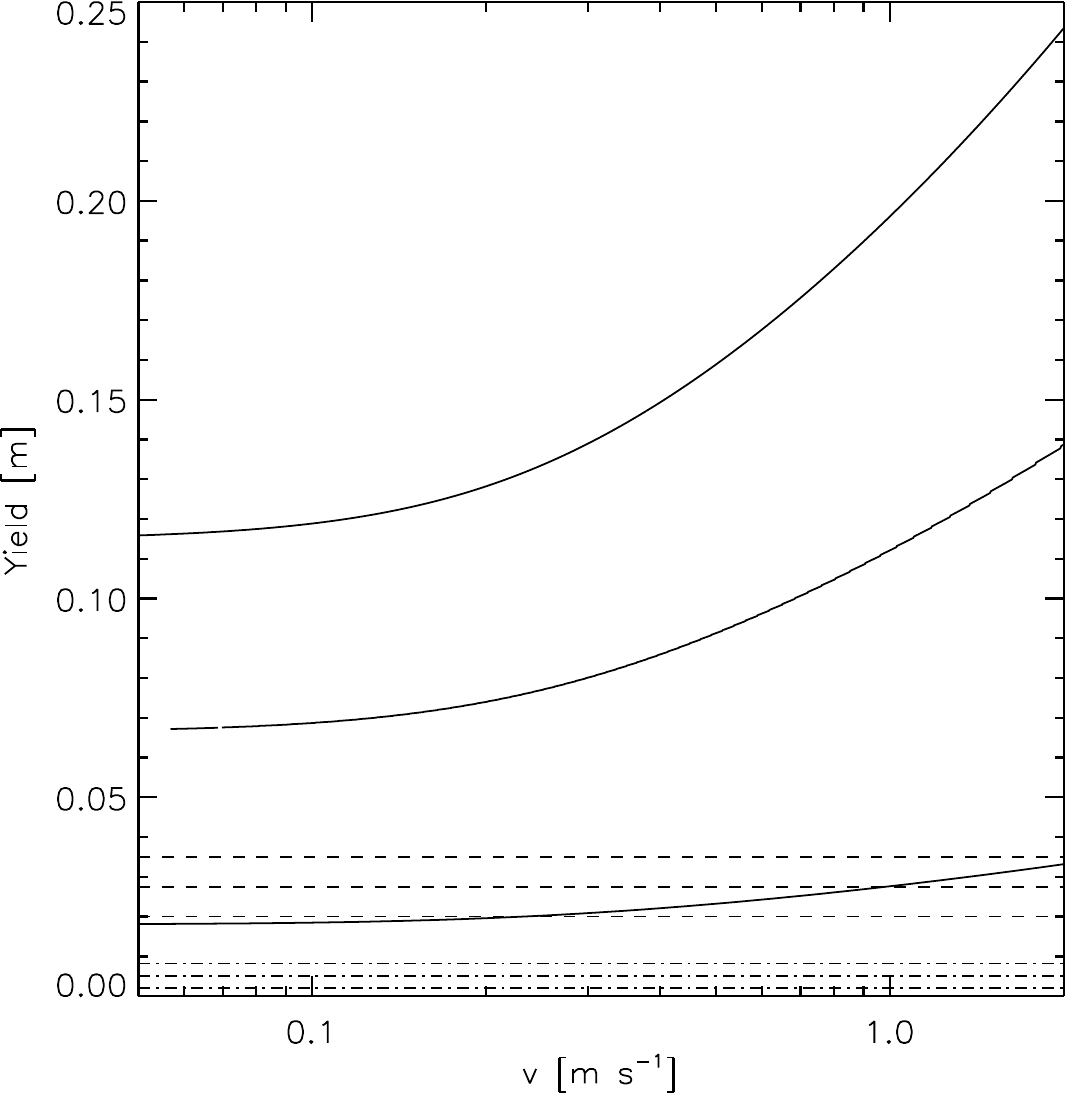}%\columnwidth
    \caption{\label{fig:67Y} The yield of the Rosetta lander (assuming a mass of 100 kg and a total landing-feet area 0.07m$^2$) whose landing gear decelerates the impact within a length of 10 cm on an agglomerate-layer material consisting of 180 $\mu$m-sized agglomerates (solid lines), 1 mm-sized agglomerates (dashed lines) 1 cm sized agglomerates (dashed-dotted lines), as a function of impact velocity. Within a group of this lines the uppermost curve is in case the surface consists of ice agglomerates, the lowermost curve in case of dust agglomerates and the center curve in case of a 50\%-50\% (in volume) mixture of dust and ice agglomerates. }
\end{figure}

 In Fig. \ref{fig:67Y}, the solid group of three lines show the yield of a spacecraft landing on regolith consisting of 180~$\mu$m-sized  aggregates as a function of impact velocity. The dashed and dash-dotted group of three lines show the impact penetration for mm-sized agglomerates and cm -sized agglomerates. Within these groups of three lines the uppermost line corresponds to ice agglomerates the center line corresponds to  a 50\%-50\% (in volume) mixture of ice and dust agglomerates and the lowermost line corresponds to dust agglomerates..

At low velocities ($v_{\mathrm{imp}}<$20~cm~s$^{-1}$) and a surface of 180~$\mu$m-sized  aggregates , the yield is dominated by the gravity of the comet. For larger velocities the impact velocity gets important and dominates the yield at velocities larger than 1~m~s$^{-1}$. In case the surface consists of agglomerates smaller than a few millimeters in diameter the yield is always dominated by the gravity of the comet.

 The penetration depth is higher for the ice agglomerates, because their higher rolling force prevented an RCP packing down to a depth of a few decimeters so that the icy material can potentially be compacted more than a pure dust-aggregate surface. However, the total penetration is less than 20 cm even for impact speeds as high as 1~m~s$^{-1}$.

%impact pressure will be insufficient to cause any compaction of the comet surface material. For slightly higher landing velocities 4.5~cm~$s^{-1}<v_{\mathrm{imp}}<$20~cm~s$^{-1}$, there will be a shallow penetration of the surface of less than 2 cm in depth for a pure dust-agglomerate surface, but still no penetration in the pure ice case, because the rolling force of the dust aggregates is lower than of the ice particles. For $v_{\mathrm{imp}}>$30~cm~s$^{-1}$, the penetration depth is higher for the ice agglomerates, because their higher rolling force prevented an RCP packing down to a depth of a few decimeters so that the icy material can potentially be compacted more than a pure dust-aggregate surface. However, the total penetration is less than 12 cm even for impact speeds as high as 1~m~s$^{-1}$. In case  p$_m$  from table \ref{table:FFP} was used for the calculation of Fig.\ref{fig:67Y} the maximum penetration depth for the ice agglomerate case was 8 cm at 1 ms$^{-1}$ landing velocity.

\section{Summary}\label{kap:COCON}
From measurements of the static compaction of regolith analog at different g-levels on-board a parabolic aircraft as well as in ground experiments, we found that the filling factor depends on the ambient hydrostatic pressure as well as on grain size and morphology. Using an analytical description for the compression curve, we developed an analytical model to predict the stratification of regolith-covered dusty and icy Solar-System bodies. We compared our findings of the stratification of the surface regolith of the Moon with Apollo 15-17 core tubes and drill cores and found only a slightly denser packing than predicted in our model. In a second step, we calculated the mechanical yield of a spacecraft resting on the surface of such a body. We found that a spacecraft with reasonable mass will never penetrate into the regolith of any Solar-System body by more than 60~cm and a comet by more than 25~cm.\\

{\bf  Acknowledgments:}
This research was supported by the Deutsches Zentrum f\"ur Luft- und Raumfahrt
under grant no. 50WM1236. We thank TU M\"unchen and the European Space Agency for
providing the parabolic flights, Novespace for yet another professional and flawless campaign onboard their parabolic aircraft, and Eike Beitz for the SEM microscopy of our 1.5$\mu$m samples.

\bigskip

\bigskip

\bigskip

\bigskip

\bigskip

\bibliographystyle{aa}
\bibliography{ms}

\appendix
\section{}
The calculation in this Section shows that the high-pressure behavior of the filling factor of dust agglomerates, as studied by \citet{Machii2013} and represented in Fig. \ref{fig:PhiH}, is independent of the size of the agglomerates. We as well as \citet{Machii2013} used in the experiments agglomerates with a filling factor of 0.35. These agglomerates are organized in RCP in a regolith with a given minimal pressure (which is depending on the agglomerate size, as shown in Fig. \ref{fig:PhiH}). Because the agglomerates can fill the void spaces in between each other under further compression, the compression curve follows the one by \citet{BS2004} for unidirectional compression. The pressure above which the grains inside the agglomerates flow is $p > p_{f}=10^5$~Pa \citet[see][]{BS2004}.

Therefore, the area of the contact circle between two agglomerates, $A_c$, depends on the force $F$ applied to the contact sphere,
\begin{eqnarray}
A_c=\frac F {p_f}.\label{eq:acf}
\end{eqnarray}
The dependency of the contact area on the indentation $h$ of the deformed agglomerates with radius $r$ is approximately given by the cut-face of a sphere
\begin{eqnarray}
A_c=(2hr-h^2) \label{eq:ach}.
\end{eqnarray}
Applying an indentation that is proportional to the radius of the agglomerates, i.e. $h/r=$~const, we get $A_c \propto r^2$. The force on the agglomerates inside the body depends on the local pressure $p$ through
\begin{eqnarray}
F=\frac{p r^2 \pi}{\Phi_{RBD}}\label{eq:fpr}
\end{eqnarray}
(see also Eq. \ref{eq:pm}). With the above condition, we see that the pressure acting at the contact area of an agglomerate is independent of the radius of the agglomerates, because both the force on the agglomerate at a given pressure inside the agglomerate and the contact area at a given force are proportional to $r^2$. This also means that a given pressure leads to a compaction of the regolith independent of the agglomerate size. This conclusion is, however, only valid for pressures high enough to compact the agglomerates, i.e. for $p > p_{f}$.

\end{document}